\documentclass[aps, pra, twocolumn, 10pt, showpacs, floatfix,superscriptaddress]{revtex4-1}
\usepackage{graphicx, verbatim, amsmath, amssymb,color,setspace,subfigure}
\usepackage{natbib}
\usepackage{epsfig}

\begin{document}


\title[Scaling in Lieb-Liniger gases]{Universal scaling of density and momentum distributions in Lieb-Liniger gases}

\author{Wei Xu}
\author{Marcos Rigol}
\affiliation{Department of Physics, The Pennsylvania State University, University Park, Pennsylvania 16802, USA}

\begin{abstract}
We present an exact numerical study of the scaling of density and momentum distribution functions of harmonically trapped one-dimensional bosons with repulsive contact interactions at zero and finite temperatures. We use path integral quantum Monte Carlo with worm updates in our calculations at finite interaction strengths, and the Bose-Fermi mapping in the Tonks-Girardeau regime. We discuss the homogeneous case and, within the local density approximation, use it to motivate the scaling in the presence of a harmonic trap. For the momentum distribution function, we pay special attention to the high momentum tails and their $k^{-4}$ asymptotic behavior.
\end{abstract}
\maketitle

\section{Introduction}
Developments in ultracold atomic gas experiments have boosted the study of many-body interaction effects in bosonic systems \cite{bloch_dalibar_08,1Dbosonrmp}. In particular, the ability to load Bose-Einstein condensates in optical lattices has provided a unique opportunity to control the effective dimensionality and interactions of Bose gases. A setup that is of interest to us here is that of a BEC loaded in a two-dimensional (2D) optical lattice, as a result of which it splits into an array of cigar-shaped Bose gases at very low temperatures \cite{1Dbosonrmp}. For a sufficiently deep 2D optical lattice and low energies, each cigar-shaped Bose gas can behave as an effective one-dimensional (1D) system \cite{greiner_bloch_01,moritz_stoferle_03} that is described by the Lieb-Liniger model \cite{liebI,liebII} (plus an additional term to account for trapping potentials). The effective 1D interaction strength is determined by the three-dimensional (3D) scattering length and the transverse confinement provided by the lattice \cite{olshaniiLL}. For sufficiently strong effective 1D interactions and sufficiently low 1D densities, these gases enter the so-called Tonks-Girardeau regime \cite{TGcont,kinoshita_wenger_05}, in which bosons behave as impenetrable particles (hard-core bosons) and can be mapped onto noninteracting fermions \cite{TGorigin}. Furthermore, if an additional (weaker, but not too weak) lattice is added along the longitudinal direction, the system can be described by the 1D version of the Bose-Hubbard model, which has been of much interest in condensed matter physics \cite{fisher_weichman_89}. Experiments in the presence of that additional lattice have allowed for the observation of the superfluid to Mott insulator transition in one dimension \cite{stoferle_moritz_04} and the lattice version of the Tonks-Girardeau regime \cite{TGlatt}.

The Lieb-Liniger model has attracted quite some attention since its introduction more than 50 years ago \cite{1Dbosonrmp,yangyang}. It is an integrable model whose exact solution, which can be obtained using the Bethe ansatz \cite{liebI}, provides insights into the universal behavior of 1D gapless systems. Its far-from-equilibrium dynamics has been recently scrutinized theoretically \cite{iyer_andrei_12,caux_konik_12,iyer_guan_13,kormos_shashi_13,nardis_wouters_14,zill_wright_15}, after experiments showed that 1D Bose gases taken far from equilibrium relax to states in which observables are not described by traditional ensembles of statistical mechanics \cite{kinoshita_wenger_06} and generalizations of the latter are needed \cite{rigol_dunjko_07,cassidy_clark_11}. 

We should stress that despite the fact that the Lieb-Liniger model is Bethe ansatz solvable, the calculation of correlation functions is extremely challenging due to the complexity of the Bethe ansatz eigenfunctions. In the ground state, for example, the evaluation of one-particle correlations (of which the momentum distribution function is the Fourier transform) has only been possible numerically within algebraic Bethe ansatz \cite{caux_calabrese_07}, and through diffusion Monte Carlo simulations \cite{astrakharchikCF,astrakharchik_giorgini_06}. At finite temperatures, they have been calculated using simulations within the stochastic gauge method \cite{deuar_sykes_09}. Remarkably, the asymptotic behavior of the momentum distribution function [$m(k)$], for large values of the momentum $k$ (associated with short range correlations), has been determined analytically [$m(k)\propto k^{-4}$] \cite{LLtail}. In the Tonks-Girardeau regime, the simplification introduced by the mapping to noninteracting fermions has enabled the analytic calculation of one-particle correlations at zero and finite temperatures \cite{lenard_64,lenard_66,vaidya_tracy_79,jimbo_80}. On the other hand, the 1D Bose-Hubbard model {\it is not} integrable, so correlations can only be computed numerically \cite{1Dbosonrmp}. However, its Tonks-Girardeau limit is integrable and can be mapped onto the so-called $XX$ chain \cite{1Dbosonrmp}, which is the isotropic limit of the $XY$ chain introduced by Lieb, Shultz, and Mattis \cite{lieb_shultz_61}, and which can be mapped onto noninteracting fermions \cite{1Dbosonrmp}. This way, it is possible to calculate analytically one-particle correlations of the Tonks-Girardeau gas in the presence of a lattice \cite{mccoy_68,vaidya_tracy_78}.

In experiments with ultracold gases, a confining potential (that is, to a good approximation, harmonic) is needed to contain the gas. In the presence of such a confining potential, as in the Hubbard model, only the limit in which bosons are impenetrable (hence, mappable to noninteracting fermions) remains integrable. In this limit, the ground-state density and momentum distribution functions, as well as one-particle correlations, have been studied in the continuum \cite{girardeau_wright_01,minguzzi_vignolo_02,lapeyre_girardeau_02,papenbrock_03,forrester_03,gangardt_04} and in the lattice \cite{rigolUni,rigolGS}. In those studies, universal power laws with the same exponents as the ones known from homogeneous systems were found. At finite temperatures, a systematic study of one-particle observables has only been reported in the lattice \cite{rigolFT}, while, in the continuum, the momentum distribution function was recently calculated in Ref.~\cite{FTmdf} for trapped systems with five particles. For finite interaction strengths, for which trapped systems are not integrable anymore, quantum Monte Carlo simulations (with a lattice discretization) have been used to compute momentum distribution functions in the continuum and compare them with experimental results at relatively weak interactions, obtaining a good agreement between the two \cite{MDFboson}.

In this work, we are interested in systematically studying the density and momentum distribution functions of the Lieb-Liniger model in a harmonic trap at zero and finite temperature. We will focus on the scaling properties of those quantities and in the high momentum tails of the momentum distribution function. The latter have remained elusive to quantum Monte Carlo simulations so far. The weight of the high momentum tail, which is known as Tan's contact, is a quantity that plays a central role in a set of universal thermodynamic relations known as Tan's relations~\cite{tanI,tanII,tanIII}. The Tan's contact has been measured in experiments with 3D Bose-Einstein condensates \cite{wild_makotyn_12}. 

In order to obtain very accurate results for all the quantities above, for arbitrary interaction strengths, here we use path integral quantum Monte Carlo \cite{pimc} for continuous systems with worm updates \cite{worm1,worm2}. We also obtain results in the Tonks-Girardeau limit using the Bose-Fermi mapping in the lattice and working at very low fillings \cite{rigolUni,rigolGS,rigolFT}. For completeness, in all cases, we also discuss the behavior of the quantities of interest in homogeneous systems, which helps motivate the scaling relations for the trapped systems.

The paper is organized as follows. In Sec.~\ref{sec:hamilQMC}, we introduce the Hamiltonian and numerical approaches used. We also discuss some of the checks done to gauge the accuracy of our calculations. In Secs.~\ref{sec:groundstate} and \ref{sec:finitetemperature}, we report a detailed study of density and momentum distribution functions, as well as one-particle correlations, of Lieb-Liniger systems at zero and finite temperatures, respectively. In both sections, we discuss results for homogeneous and harmonically trapped systems. Finally, our conclusions are presented in Sec.~\ref{sec:conclusion}.

\section{Hamiltonian and numerical approaches} \label{sec:hamilQMC}
We consider 1D bosons with repulsive contact interactions in the presence of an external harmonic trap. The Hamiltonian can be written as
\begin{equation}
 \mathcal{H}=\sum_j\left[-\frac{\hbar^2}{2m}\frac{\partial^2}{\partial x_j^2}+V(x_j)\right]+g\sum_{j<l}\delta(x_j-x_l),
\end{equation}
where $m$ is the mass of the bosons, $g$ is the strength of the contact interaction, $V(x_j)=m\omega^2 x_j^2/2$, and $\omega$ is the frequency of the harmonic trap. The strength of the effective 1D contact interaction is usually written as $g=-2\hbar^2/ma_\mathrm{1D}$, where $a_\mathrm{1D}=-a_\perp(a_\perp/a_s-C)$ is the 1D scattering length, $a_s$ is the (3D) $s$-wave scattering length, $a_\perp$ is the length of the transverse confinement, and $C=1.0326$~\cite{olshaniiLL}. Another parameter frequently used to describe the interaction strength is $c=mg/\hbar^2$. In the absence of the trap, this Hamiltonian reduces to the Lieb-Liniger model, which essentially has only one tunable parameter $\gamma=c/\rho$, where $\rho$ is the density.

Here, in order to simulate the Hamiltonian above in the continuum, we use a path integral quantum Monte Carlo method with worm updates (from now on referred to as the worm algorithm) \cite{worm1,worm2}. The configuration space in our simulation is built with discrete imaginary time world lines in a continuous position space. The worm algorithm operates both with diagonal and off-diagonal configurations by introducing one additional open world line in the configuration space. This makes it very efficient in simulating both diagonal observables (such as the energy, the density distribution, and the superfluid 
fraction) and off-diagonal observables (such as two-point one-particle correlation functions). In our calculations, the action for contact interactions is approximated by the pair-product action, see Appendix~\ref{app:PPU}, and we always work in the grand canonical ensemble.

\begin{figure}[!t]
 \includegraphics[width=0.75\linewidth]{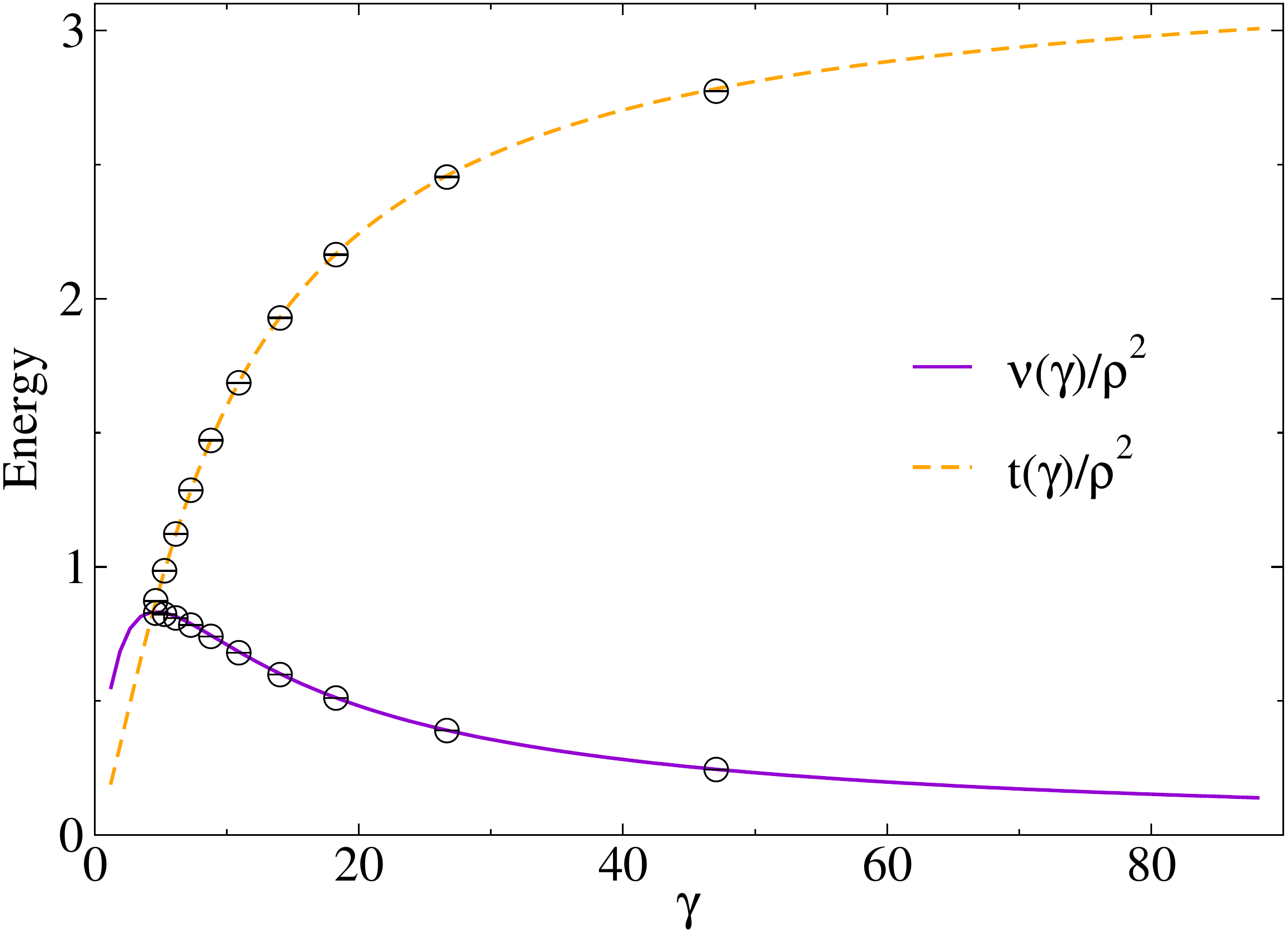} 
 \caption{(Color online) Worm algorithm results for the ground-state kinetic and interaction energies (see text) of the Lieb-Liniger model (circles) compared with the exact analytical results (lines) \cite{liebI}. The statistical errors (shown within the circles) can be seen to be much smaller than the circle sizes. The average number of particles in the system is $N_b\approx20$.}
 \label{fig:liebenergy}
\end{figure}

In Fig.~\ref{fig:liebenergy}, we compare results for the ground state kinetic [$K(\gamma)=\hbar^2 t(\gamma)/{2m}$] and interaction  [$U(\gamma)=\hbar^2 \nu(\gamma)/{2m}$] energies from the worm algorithm simulations (for a finite system with $N_b\approx20$ bosons) with the exact analytic results in the thermodynamic limit~\cite{liebI}. Average energies are obtained using the thermodynamic estimators reported in Appendix~\ref{app:E_est} (see also Ref.~\cite{pimc}). Figure~\ref{fig:liebenergy} shows that, already for systems with $N_b\approx20$, finite-size effects are negligible, and that the worm algorithm provides a very accurate estimation of observables in diagonal configuration space.

The observables on which we focus in this work, i.e., the density and momentum distribution functions, are calculated from the one-particle density matrix 
\begin{equation}
 g(x,y)=\langle\Psi^\dag(x)\Psi(y)\rangle,
\end{equation}
where $\Psi^\dag(x)$ [$\Psi(x)$] is the bosonic field creation (annihilation) operator at position $x$. $g(x,y)$ is an example of an off-diagonal observable that can be efficiently calculated using the worm algorithm. 

For improved statistics, in our calculations, we average $g(x,y)$ over all possible translations ($0<x,y<L$) so that it becomes a function of $r=|x-y|$. For $r>0$,
\begin{equation}
 g(r)=\frac1L\left[\int_0^Lg(x,x+r)dx+\int_0^Lg(y+r,y)dy\right],
\end{equation}
while, for $r=0$,
\begin{equation}
 g(0)=\frac1L\int_0^Lg(x,x)dx.
\end{equation}
The Fourier transform of $g(r)$ gives the momentum distribution function
\begin{eqnarray}
 m(k)&\equiv&\frac{1}{2\pi}\int_0^L dx\int_0^L dy \, e^{ik(x-y)}g(x,y)\nonumber\\&=&
 \frac{L}{2\pi}\int_0^L \cos(kr)g(r)dr.
 \label{eq:mdf_cont}
\end{eqnarray}

\begin{figure}[!t]
 \centering
 \includegraphics[width=0.85\linewidth]{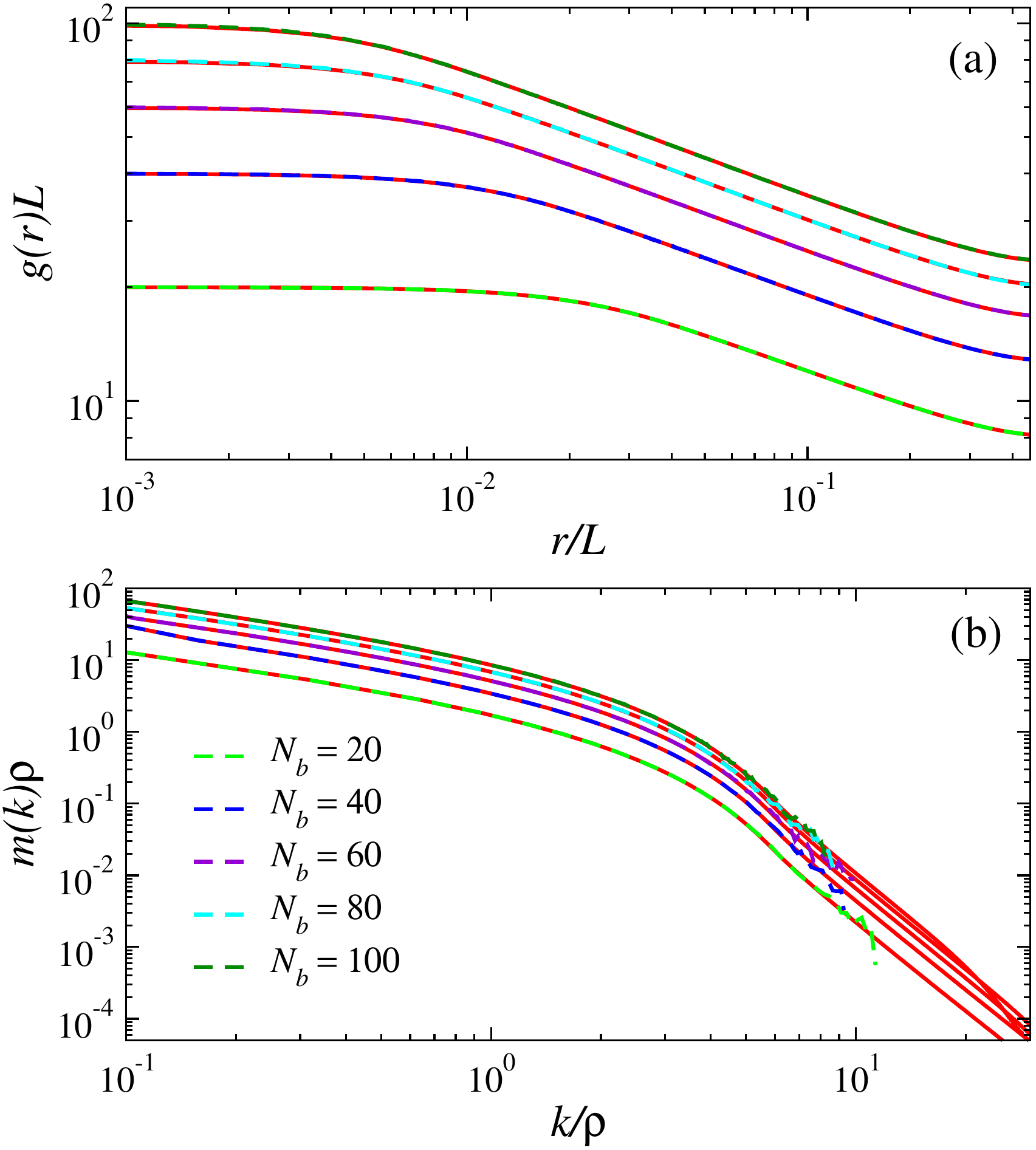} 
 \caption{(Color online) (a) Ground-state one-particle correlations and (b) momentum distribution functions of bosons for $\gamma=8.33$. Results are reported for systems with the same density but different average number of particles. Dashed lines are the results obtained using the worm algorithm and solid lines the results obtained in numerical calculations via algebraic Bethe ansatz~\cite{caux_calabrese_07}.}
 \label{fig:LL_JCaux}
\end{figure}

In Fig.~\ref{fig:LL_JCaux}, we show results for $g(r)$ [Fig.~\ref{fig:LL_JCaux}(a)] and $m(k)$ [Fig.~\ref{fig:LL_JCaux}(b)] for the ground state of systems with $\gamma=8.33$, the same density, and different average number of particles, and compare them with numerical results obtained via algebraic Bethe ansatz~\cite{caux_calabrese_07}. The results for $g(r)$ in both approaches are essentially indistinguishable in Fig.~\ref{fig:LL_JCaux}(a). The same is true for $m(k)$ in Fig.~\ref{fig:LL_JCaux}(b), except at the highest momenta. In the latter regime, $m(k)$ becomes very small and the statistical errors in the quantum Monte Carlo simulation become of the same order. This is why it is a challenge to study the asymptotic behavior of $m(k)$ at high momenta. Experiments will suffer from exactly the same limitation as $m(k)$ will at some point become of the same order of the experimental noise. The results in Fig.~\ref{fig:LL_JCaux} make apparent the accuracy of the worm algorithm for computing off-diagonal one-particle observables, which are the focus of this work. Once translational invariance is broken, no analytical solution is available. To gauge the accuracy of the worm algorithm in that case we focus on the Tonks-Girardeau limit. 

In that limit, the one-particle density matrix (and, hence, the momentum distribution function) in a lattice can be calculated exactly via the Jordan-Wigner transformation and making use of properties of Slater determinants, both at zero \cite{rigolUni,rigolGS} and finite temperature \cite{rigolFT}. As we explain below, using this approach in the low-density limit in the lattice, one can efficiently compute one-particle properties in the continuum \cite{rigolGS}. 

The Tonks-Girardeau Hamiltonian in the lattice can be written as
\begin{equation}
 \mathcal{H}_\mathrm{TG}=-t\sum_i(b_i^\dag b^{}_{i+1}+\text{H.c.})+V\sum_i i^2n_i,
 \label{Eq:H_TG}
\end{equation}
where $b_i^\dag$ ($b_i^{}$) is the creation (annihilation) operator of a hard-core boson at site $i$, $n_i=b_i^\dag b_i^{}$ is the site occupation operator, $t$ is the hopping amplitude, and $V$ sets the strength of the harmonic trap. This Hamiltonian is obtained from the Bose-Hubbard Hamiltonian by taking the limit in which the onsite repulsion $U\rightarrow\infty$ \cite{1Dbosonrmp}. 

The hard-core boson creation and annihilation operators satisfy standard bosonic commutation relations with the constraint $(b_i^\dag)^2=b_i^2=0$. Mapping the Tonks-Girardeau Hamiltonian \eqref{Eq:H_TG} onto a spin-1/2 chain \cite{rey_satija_06,he_rigol_11}, and then the spin-1/2 chain onto noninteracting fermions in 1D via the Jordan-Wigner transition, one gets
\begin{equation}\label{Eq:H_F}
 \mathcal{H}_F=-t\sum_i(f_i^\dag f^{}_{i+1}+\text{H.c.})+V\sum_i i^2 n^f_i,
\end{equation}
where $f_i^\dag$ ($f_i^{}$) is the creation (annihilation) operator of a spinless fermion at site $i$, and $n^f_i=f_i^\dag f_i^{}$ is the fermionic site occupation operator. The fermionic Hamiltonian \eqref{Eq:H_F} can be straightforwardly diagonalized. One-particle bosonic correlations can then be obtained using properties of Slater determinants as discussed in Refs.~\cite{rigolUni,rigolGS,rigolFT}. We should stress that, within this lattice approach, the ground-state calculations are done in the canonical ensemble \cite{rigolUni,rigolGS}, while finite-temperature ones are done in the grand-canonical one \cite{rigolFT}.

To establish the relation between the parameters in the lattice Hamiltonian and in the continuum, we notice that the single-particle energy spectrum of the spinless fermion model without a trap is $\varepsilon_k=-2t\cos(ka)$, where $a$ is the lattice spacing. In the zero-density limit, the Fermi momentum $k_F\rightarrow 0$. As a result, the energy spectrum becomes quadratic in $k$ as, up to a constant, we can write $\varepsilon_k=ta^2k^2$. Thus, in the zero-density limit, the lattice model reduces to a continuum model with the effective mass given by $m=\hbar^2/2ta^2$. In this limit, according to Eq.~\eqref{Eq:H_TG}, the external trapping potential is
\begin{equation}
 V(x)=V(x/a)^2 .
\end{equation}
The trapping frequency is then given by $\omega=2\sqrt{tV}/\hbar$, and so the harmonic oscillator length $a_\mathrm{HO}=\sqrt{\hbar/m\omega}$ can be written as $a_\mathrm{HO}/a=(t/V)^{1/4}$. Finally, in finite-temperature calculations in the lattice, the lattice temperature $T_t$ is usually defined as $t\,T_t=k_BT$, i.e., it is given in units of $t$. The temperature in units of trapping frequency in the continuum is related to the lattice temperature by the expression $k_BT/\hbar\omega=T_t\sqrt{t/V}/2$.

\begin{figure}[!t]
 \centering
 \includegraphics[width=0.95\linewidth]{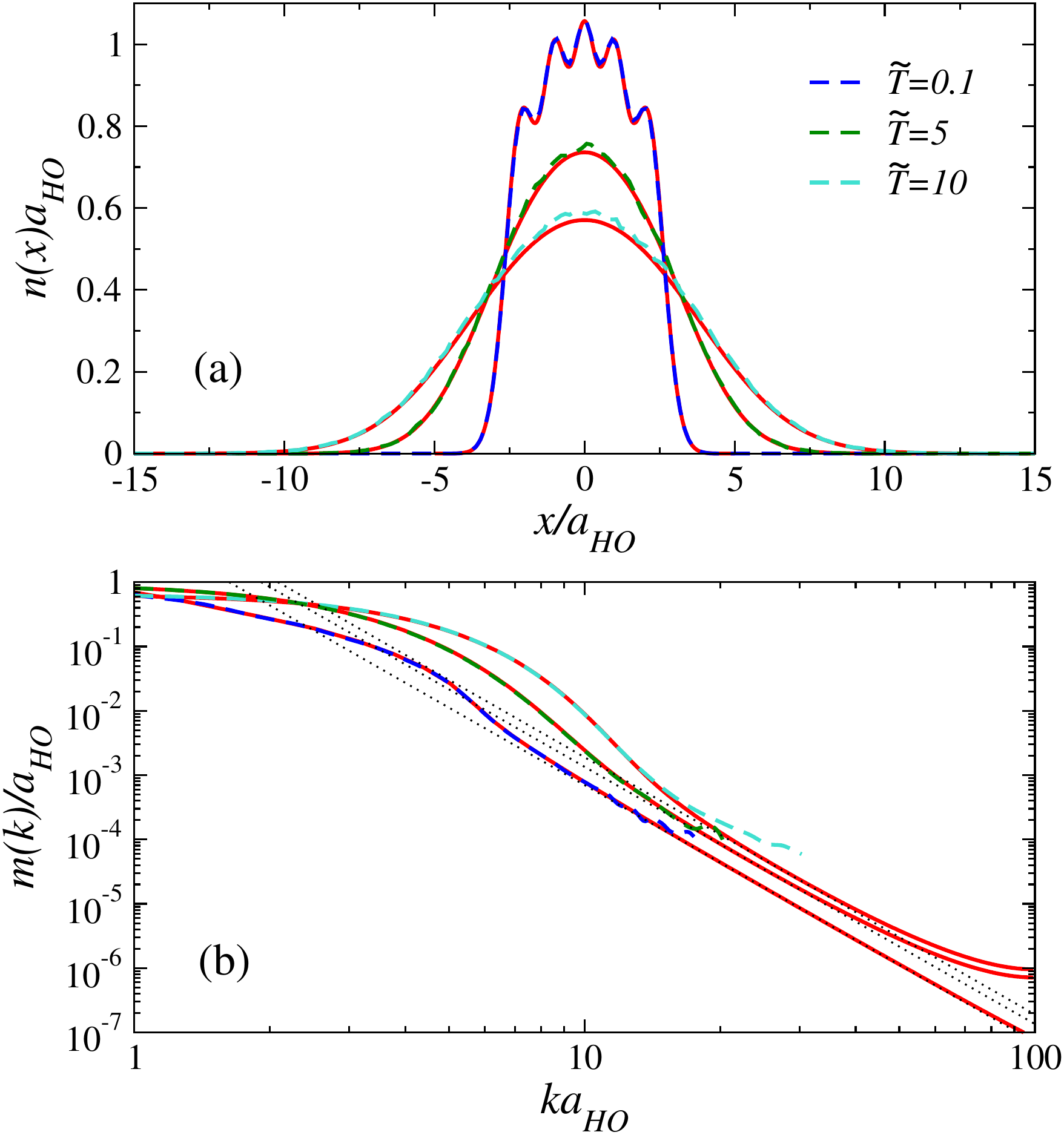} 
 \caption{(Color online) Density profiles (a) and momentum distribution functions (b) of the Tonks-Girardeau gas as calculated using the lattice approach (red solid lines) and in the continuum using the worm algorithm (dashed lines). The average number of particles is 5. Results are presented for different temperatures: $\tilde{T}\equiv k_BT/\hbar\omega$=0.1, 5, 10. Thin dotted lines indicate the $k^{-4}$ asymptotic behavior of the momentum tails. The statistical errors in the worm algorithm are of the order of the fluctuations seen in the results. They are not reported for clarity.}
 \label{fig:ContvsLatt_Nb5}
\end{figure}

In the lattice, the momentum distribution function is defined as the discrete Fourier transform of the one-particle density matrix
\begin{equation}
 m_k=\frac1N\sum_{jl}e^{ika(j-l)}b_i^\dag b^{}_{j},
 \label{eq:mdf_latt}
\end{equation}
where $N$ is the number of lattice sites. By comparing Eq.~\eqref{eq:mdf_cont} and Eq.~\eqref{eq:mdf_latt}, one can see that the momentum distribution function in the continuum and in the lattice are related by the expression $m_k=m(k)\delta k$, where $\delta k=2\pi/L$ (with $L=Na$) is the discretization of $k$ in the lattice model.

With the relations discussed so far at hand, we are ready to compare results for density and momentum distribution functions in the continuum and the lattice, which, as mentioned before, help us gauge the accuracy of the worm algorithm for diagonal and off-diagonal observables in the absence of translational invariance. In order to approach the Tonks-Girardeau limit in the continuum model, we choose a very small value of the 1D scattering length $a_\mathrm{1D}/a_\mathrm{HO}=5.66\times10^{-3}$.

\begin{figure}[!t]
 \centering
 \includegraphics[width=0.9\linewidth]{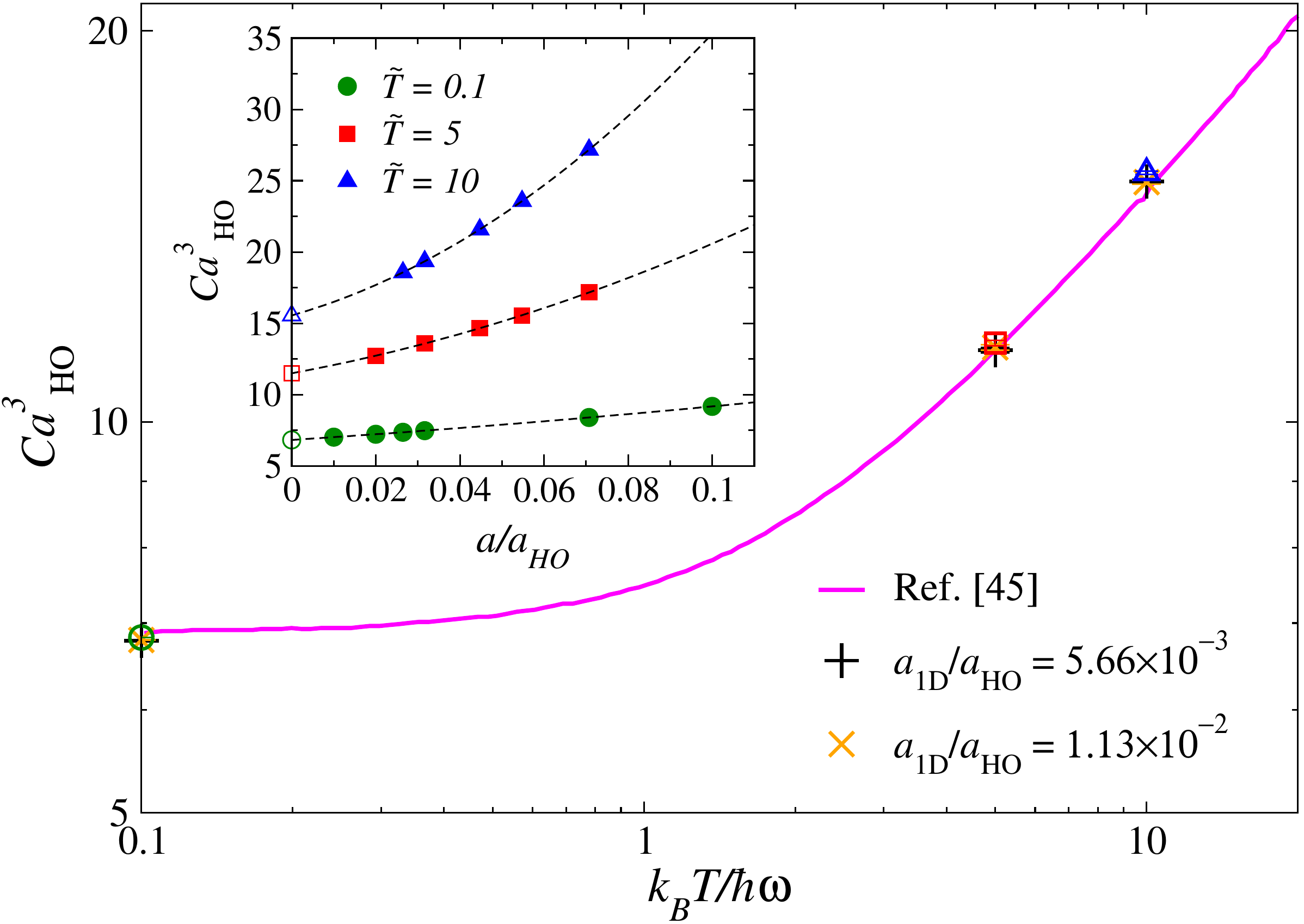} 
 \caption{(Color online) Tan's contact for the trapped systems in Fig.~\ref{fig:ContvsLatt_Nb5} compared with results in Ref.~\cite{FTmdf}. The contact from the worm algorithm simulation is obtained through the interaction energy using the expression $\mathcal{C}a^3_\mathrm{HO}=-a^3_\mathrm{HO}N_b\nu(\gamma)/(\pi a_\mathrm{1D})$ [derived from Eq.~\eqref{eq:TanC}], where $\nu(\gamma)$ is the average interaction energy per particle [in units of $\hbar^2/(2m)$]. We calculated $\mathcal{C}a^3_\mathrm{HO}$ for two small values of $a_\text{1D}$ obtaining results that agree with the ones in Ref.~\cite{FTmdf}. The contact from the lattice calculation is obtained via a linear fit of the high momentum tails, and then extrapolated to the zero density limit via a quadratic fit to obtain the result for zero $a/a_\mathrm{HO}$ (see inset).}
 \label{fig:c_tgtemp}
\end{figure}

The scaled density [$n(x)=g(x,x)$] and momentum distribution functions obtained in the lattice and the continuum are shown in Fig.~\ref{fig:ContvsLatt_Nb5} at various temperatures ($\tilde{T}\equiv k_BT/\hbar\omega$). The agreement between the results of both approaches is almost perfect at the lowest temperatures $\tilde T=0.1$ and 5 (for $\tilde T=0.1$ the system is essentially in the ground state). As the temperature increases, lattice effects become important and lead to visible differences when the lattice results are compared with the results in the continuum [see the discussion after Eq.~\eqref{eq:TanC}]. In the momentum distribution function [Fig.~\ref{fig:ContvsLatt_Nb5}(b)], the worm algorithm at $\tilde T=0.1$ and 5 already captures part of the $k^{-4}$ momentum tails. For higher temperatures, such as $\tilde T=10$ in Fig.~\ref{fig:ContvsLatt_Nb5}(b), the $k^{-4}$ regime is not visible in the worm algorithm calculation. This is understandable because, as the lattice calculation clearly shows, the $k^{-4}$ tail starts at higher values of the scaled momentum [smaller values of the scaled $m(k)$] as the temperature increases.

We have studied the weight $\mathcal{C}$ (Tan's contact) of the $k^{-4}$ momentum tails [$m(k)=\mathcal{C}\,k^{-4}$] using both methods. In the lattice approach, $\mathcal C$ is computed directly via linear fits of the momentum tails. In the worm algorithm calculation, the contact $\mathcal{C}$ is calculated via the interaction energy estimator, using the relation \cite{FTmdf}
\begin{equation}
 \mathcal{C}=\frac{gm^2}{\pi\hbar^4}\langle\mathcal{H}_\mathrm{int}\rangle .
\label{eq:TanC}
\end{equation}
The results are shown in Fig.~\ref{fig:c_tgtemp}. The values of the contact obtained using the worm algorithm for two small values of the scattering length are consistent with the results in Ref.~\cite{FTmdf}. This indicates that such small values of $a_\text{1D}$ allow us to obtain results in the Tonks-Girardeau limit. Using the lattice approach, the contact in the continuum is calculated by taking the low density limit by means of a quadratic fit (see the inset in Fig.~\ref{fig:c_tgtemp}). The results of such a fitting procedure are in agreement with those obtained using the worm algorithm and Ref.~\cite{FTmdf}, as shown in the main panel in Fig.~\ref{fig:c_tgtemp}. Note that for the ground state calculation, lattice effects are negligible at the lowest densities studied.

\section{Ground-State} \label{sec:groundstate}

\subsection{Homogeneous Systems} \label{sec:homT0}

\subsubsection{Tonks-Girardeau limit}\label{sec:homT0TG}

\begin{figure}[!t]
 \centering
 \includegraphics[width=1.0\linewidth]{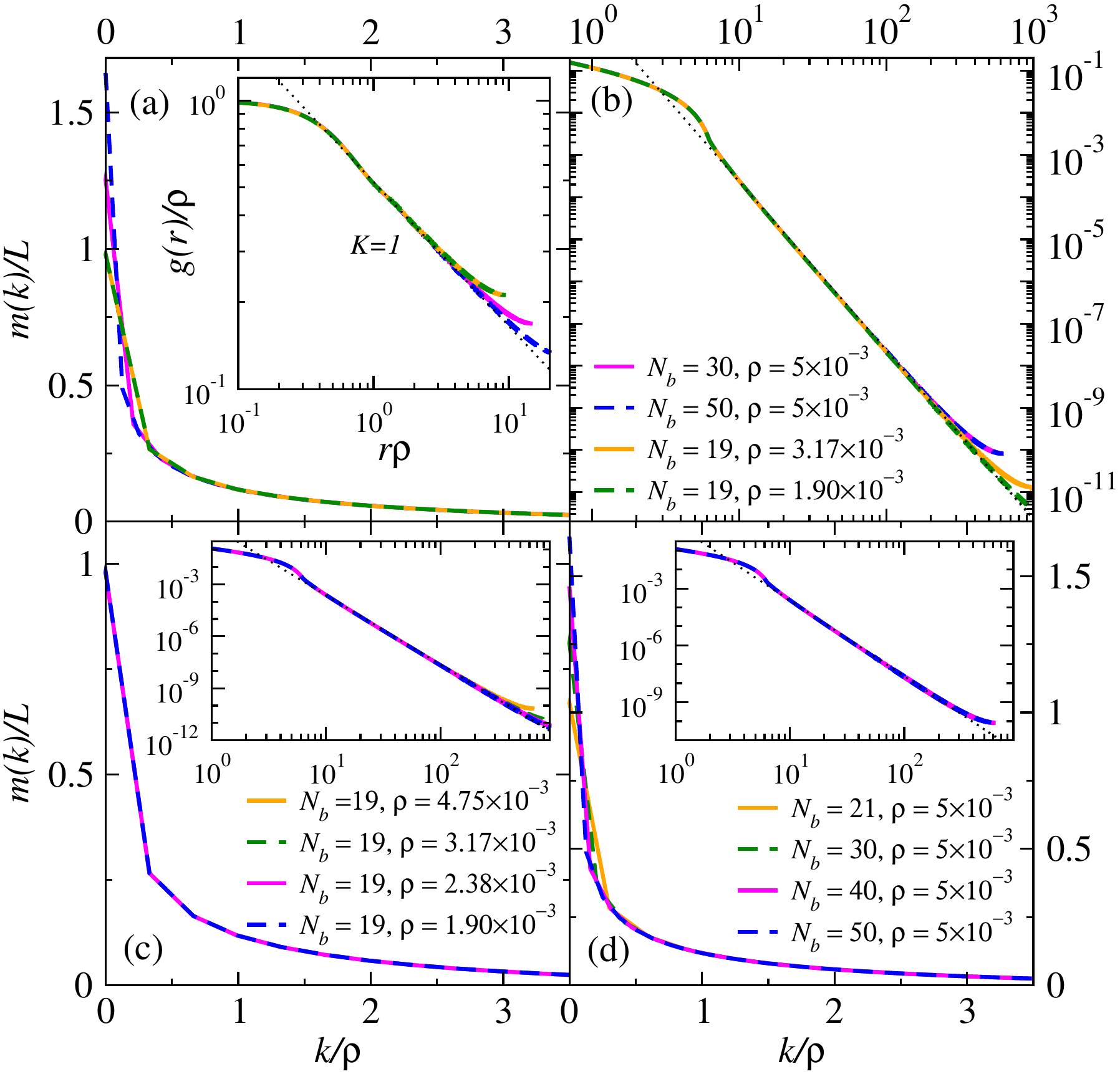}
 \caption{(Color online) Scaled momentum distribution functions $m(k)/L$ in the Tonks-Girardeau limit as a function of the scaled momentum $k/\rho$. (a) Systems with different number of particles and densities, (b) same as (a) but in log-log scale, (c) systems with the same number of particles and different densities, (d) systems with different number of particles and the same density. Insets: (a) scaled average one-particle density matrix, (c) and (d) same as main panels but in log-log scale. Thin dotted lines indicate the asymptotic behavior discussed in the text. Here, and throughout this work, $\rho$ in the lattice is given in units of the inverse lattice spacing $a$.}
 \label{fig:mdist_TG}
\end{figure}

In the Tonks-Girardeau limit, Tan's contact allows for a simple analytic expression that unveils how to scale the momentum distribution function (one-particle correlations) to achieve data collapse for high momenta (small distances). As indicated in Eq.~\eqref{eq:TanC}, $\mathcal{C}$ is linearly related to the interaction energy. In the language of Ref.~\cite{liebI}, 
\begin{equation}
 \mathcal{C}=\frac{N_b\rho\gamma\nu(\gamma)}{2\pi},
\label{eq:LLTanC}
\end{equation}
where $\nu(\gamma)=\rho^2\gamma de(\gamma)/d\gamma$, and $e(\gamma)$ is a monotonically increasing function of $\gamma$ that saturates at $\pi^2/3$ when $\gamma\rightarrow\infty$ (for its definition, see Ref.~\cite{liebI}). For large values of $\gamma$, $\nu$ has the following asymptotic form: $\nu(\gamma)=4\rho^2e(\gamma)/(\gamma+2)$. Thus, in the limit $\gamma\rightarrow\infty$, Tan's contact reduces to $\mathcal{C}=2\pi L\rho^4/3$. Hence, $\tilde{\mathcal{C}}\equiv\mathcal{C}/L\rho^4=2\pi/3$. This means that if one plots $m(k)/L$ vs $k/\rho$, no matter the number of particles and the system size, the curves must collapse for high values of $k$. From Eq.~\eqref{eq:mdf_cont}, it then follows that the universal behavior of $m(k)$ for high values of $k$ implies that, if one plots $g(r)/\rho$ vs $\rho r$, the curves must collapse for small values of $r$. Results for $m(k)$ and $g(r)$, obtained using the {\it lattice approach} discussed in the previous section, are reported in Fig.~\ref{fig:mdist_TG}

In Fig.~\ref{fig:mdist_TG}(a), we plot $m(k)/L$ vs $k/\rho$ for systems with different number of particles and densities. Figure~\ref{fig:mdist_TG}(b) reports the same results but in a log-log scale. Both panels make apparent that, for intermediate and high momentum, all curves collapse onto a universal result. In the $k\rightarrow 0$ limit, $m(k=0)/L=m_{k=0}\propto\sqrt{N_b}$ because of the existence of quasi-long-range order \cite{1Dbosonrmp} and, as such, only curves with the same number of particles can collapse for low values of $k$. This can be better seen in Fig.~\ref{fig:mdist_TG}(c), where, for the same number of particles, all momentum distribution curves collapse for all momenta. Differences can only be seen at the highest momenta because of finite-density effects in the lattice. On the other hand, as shown in Fig.~\ref{fig:mdist_TG}(d), plotting results for the same density and different number of particles, again produce curves that collapse at intermediate and high momenta but differ for the lowest values of $k$. For systems with larger number of particles, the universal behavior can be seen to start at lower values of $k$. Consequently, irrespectively of the density in thermodynamically large system sizes, one expects the momentum distribution functions to become universal starting at infinitesimal values of $k$.

In the inset in Fig.~\ref{fig:mdist_TG}(a), we plot results for $g(r)/\rho$ vs $\rho r$ for systems with different number of particles and densities. The plots can be seen collapse onto universal results, in which, at long distance, $g(r)\propto1/\sqrt{r}$ \cite{1Dbosonrmp}. The latter behavior can be understood within the bosonization approach \cite{bosonizing}, which predicts that $g(r)\propto1/r^{1/(2K)}$ at long distances, where $K$ is the so called Tomonaga-Luttinger parameter. In the Tonks-Girardeau limit, $K=1$.

\subsubsection{Finite interaction strength}\label{sec:homT0LL}

\begin{figure}[!b]
 \centering
 \includegraphics[width=1.0\linewidth]{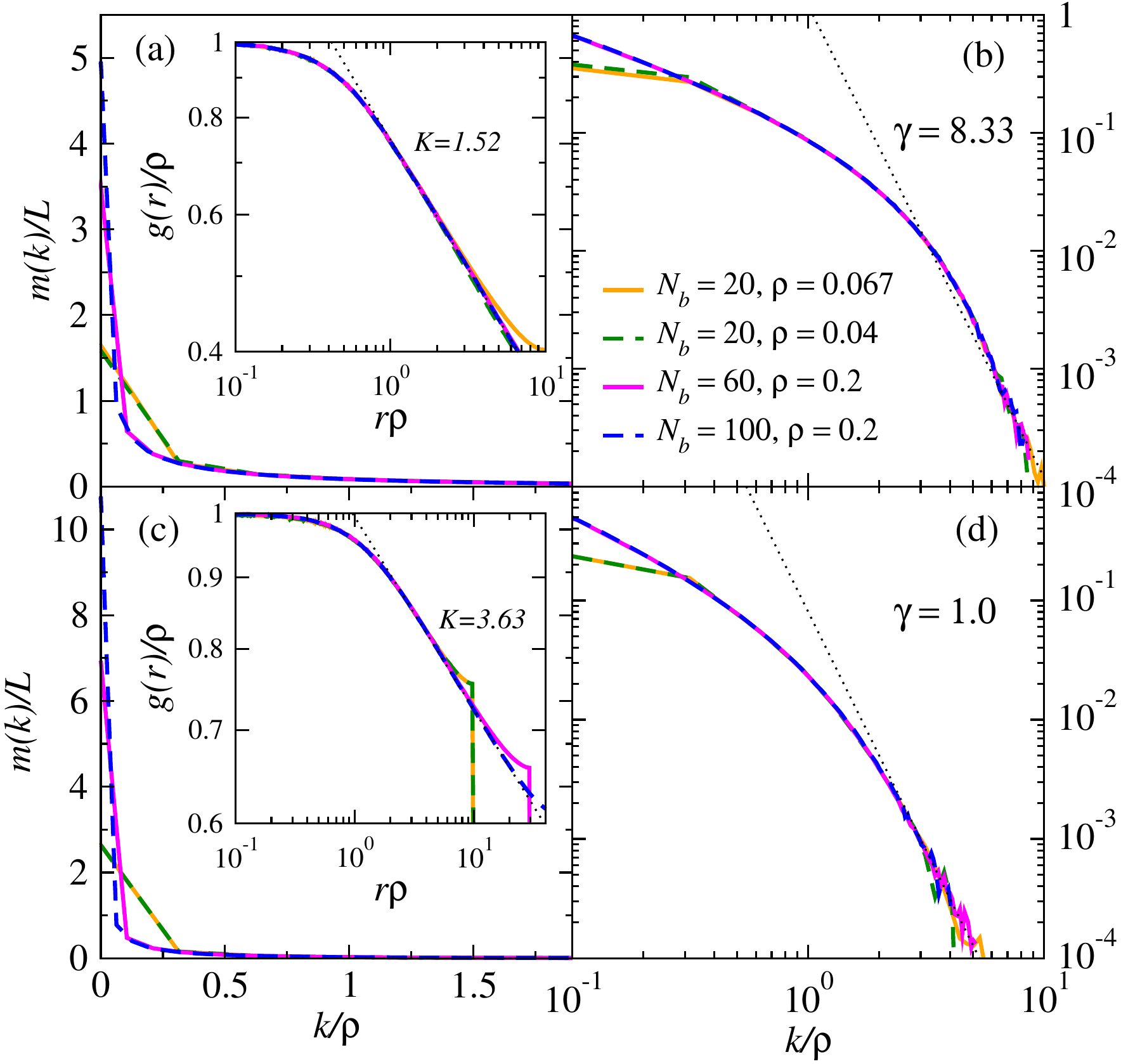}
 \caption{(Color online) Scaled momentum distribution functions $m(k)/L$ vs the scaled momentum $k/\rho$ for finite values of $\gamma$, plotted in [(a) and (c)] a linear scale and in [(b) and (d)] a log-log scale. Results are reported for [(a) and (b)] $\gamma=8.33$ and [(c) and (d)] $\gamma=1.0$. Insets: $g(r)/\rho$ vs $\rho r$ for the same systems as in the main panels. In all plots, thin dotted lines indicate the asymptotic behavior discussed in the text. A linear fit of the long distance behavior of the curves in the insets indicate that the Luttinger parameter is $K\approx1.52$ for $\gamma=8.33$ and $K\approx3.63$ for $\gamma=1.0$. Here, and throughout this work, $\rho$ in the continuum is reported in an arbitrary unit.}
 \label{fig:mdist_Gama}
\end{figure}

For finite interaction strengths, particles can penetrate through one another and the systems tend to have lower kinetic energies (less particles in the high momentum tails) than in the Tonks-Girardeau limit. As a result, Tan's contact decreases with decreasing $\gamma$. From Eq.~\eqref{eq:LLTanC}, $\mathcal{C}$ can be written as $\mathcal{C}=L\gamma^2e'(\gamma)\rho^4/(2\pi)$. The scaled contact, $\tilde{\mathcal{C}}\equiv\mathcal{C}/L\rho^4=\gamma^2e'(\gamma)/(2\pi)$, is now {\it only} a function of $\gamma$. From this result it follows that, so long as $\gamma$ is kept constant (no matter the system size and the number of particles in the system), the curves for $m(k)/L$ vs $k/\rho$ will be universal for large values of $k$. This is the same scaling discussed for the Tonks-Girardeau regime, in which $\gamma$ is constant ($\gamma=\infty$). Similarly, the curves for $g(r)/\rho$ vs $\rho r$ must be universal at short distances. Results for $m(k)$ and $g(r)$, obtained using the {\it worm algorithm}, are reported in Fig.~\ref{fig:mdist_Gama}.

In the main panels in Fig.~\ref{fig:mdist_Gama}, we plot the momentum distribution for $\gamma=8.3$ [Fig.~\ref{fig:mdist_Gama}(a) and \ref{fig:mdist_Gama}(b)] and for $\gamma=1.0$ [Fig.~\ref{fig:mdist_Gama}(c) and \ref{fig:mdist_Gama}(d)], in systems with different densities and average number of particles. Consistent with the previous discussion, the momentum distributions collapse for intermediate and high momentum. More importantly for experiments, with decreasing $\gamma$, the onset of the $k^{-4}$ behavior becomes less sharp (see also Fig.~\ref{fig:mdist_TG}, and Fig.~5 in Ref.~\cite{caux_calabrese_07}). This makes the experimental identification of the $k^{-4}$ tails increasingly difficult as one departs from the Tonks-Girardeau limit. 

As $k\rightarrow 0$, the collapse obtained for intermediate and high values of $k$ breaks down because of quasi-long-rage order. For finite values of $\gamma$, more particles have low momenta than in the Tonks-Girardeau limit. In particular, within the bosonization approach, the zero momentum occupation is predicted to be $m(k=0)/L\propto N_b^{1-1/(2K)}$, with $K\geq1$ \cite{bosonizing}. The insets in Figs.~\ref{fig:mdist_TG}(a) and \ref{fig:mdist_TG}(c) depict $g(r)/\rho$ vs $\rho r$ for the same systems for which momentum distributions are shown in the main panels. The curves can be seen to collapse at short distances as expected. At long distances, $g(r)\propto1/r^{1/(2K)}$ as predicted by the bosonization approach, with $K\approx1.52$ for $\gamma=8.3$ and $K\approx3.63$ for $\gamma=1.0$.

\subsection{Trapped Systems}

\subsubsection{Tonks-Girardeau limit}

In order to generalize the scaling discussed in Sec.~\ref{sec:homT0TG} to harmonically trapped systems, we need to find the replacement for $\rho$ and $L$ in the presence of a trap. Let us start by discussing how to scale density profiles in a harmonic trap. Since the density profiles of impenetrable bosons are the same as those of the non-interacting spinless fermions to which they can be mapped \cite{1Dbosonrmp}, we can focus on the noninteracting spinless fermions. The density profiles of the latter systems have been studied in detail in the past \cite{kolomeisky_Newman_00,vignolo_minguzzi_00,brack_zyl_01}. For a homogeneous Fermi system, the density is a linear function of the Fermi momentum $k_F$, $\rho=k_F/\pi$. The chemical potential $\mu$ in such systems is the Fermi energy $E_F=\hbar^2k_F^2/(2m)$, so $\rho=\sqrt{2m\mu}/(\hbar\pi)$.

In a harmonic trap, the local chemical potential changes according to the relation
\begin{equation}
 \mu(x)=\mu_0-V(x),
 \label{eq:eff_CP}
\end{equation}
where $\mu_0$ is the chemical potential at the center of the trap. Within the local density approximation (LDA), the density at each position in the trap is solely determined by the local chemical potential, through the relation obtained for homogeneous systems. Hence
\begin{equation}
 n(x)=\frac{1}{\pi}\left(\tilde{\mu}-\frac{x^2}{a^4_\mathrm{HO}}\right)^{\frac12},
 \label{eq:LDA_TG}
\end{equation}
where $\tilde{\mu}=2m\mu_0/\hbar^2$. The integral of Eq.~\eqref{eq:LDA_TG} over the entire space is the number of particles $N_b$. It provides the following relation between $\tilde{\mu}$ and $N_b$, $\tilde{\mu}=2N_b/a^2_\mathrm{HO}$. The density in the center of the trap is then $\tilde{\rho}=\sqrt{2N_b}/\pi a_\mathrm{HO}$, and the position in the trap at which the density vanishes is $\tilde{L}=\sqrt{2N_b}a_\mathrm{HO}$. Using these quantities, the density profile in the trap can be written as
\begin{equation}
 n(x)=\tilde{\rho}\left[1-\left(\frac{x}{\tilde{L}}\right)^2\right]^{\frac12} .
\end{equation}
By defining the scaled density $\tilde{n}(x)=n(x)/\tilde{\rho}$ and the scaled position $\tilde{x}=x/\tilde{L}$, one obtains
\begin{equation}\label{eq:scalHCB2}
 \tilde{n}(\tilde{x})=(1-\tilde{x}^2)^{\frac12} .
\end{equation}

Since the momentum distribution function at high momenta is determined by the average short-distance one-particle correlations, from the LDA and Eq.~\eqref{eq:scalHCB2} one can advance that, using the scaling relations for homogeneous systems with $\rho\rightarrow\tilde{\rho}$ and $L\rightarrow\tilde{L}$, there will be data collapse for the momentum distribution function at high momenta.

\begin{figure}[!t]
 \centering
 \includegraphics[width=1.0\linewidth]{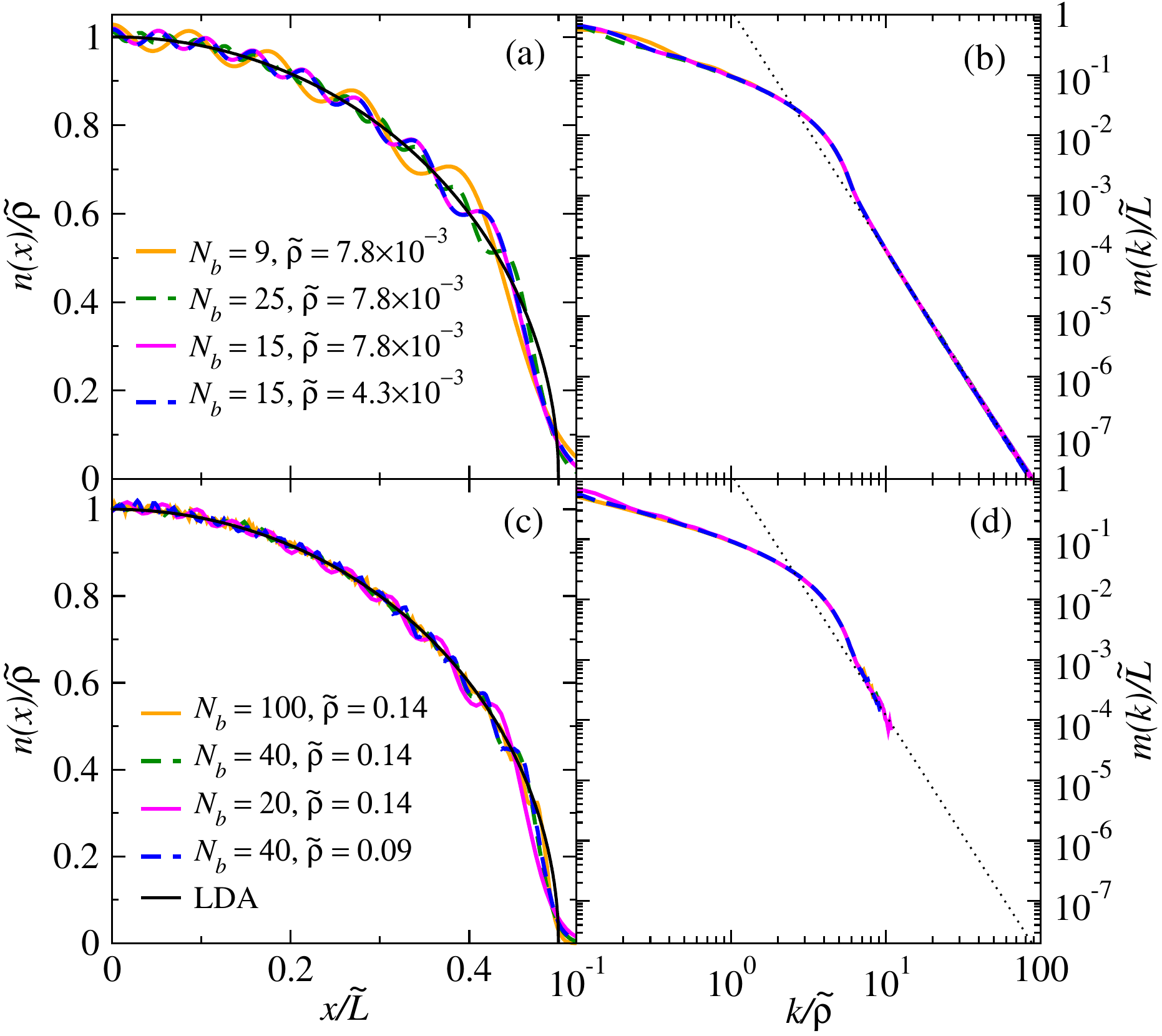}
 \caption{(Color online) [(a) and (c)] Universal scaling of density and [(b) and (d)] momentum distribution functions in the Tonks-Girardeau limit. We report results from lattice calculations for [(a) and (b)] up to 25 bosons, and from the worm algorithm for [(c) and (d)] up to 100 bosons. In (a) and (c), we also report the LDA prediction for the density profiles. In (b) and (d), thin dotted line depict $k^{-4}$ behavior.}
 \label{fig:mdistH_TG}
\end{figure}

In Fig.~\ref{fig:mdistH_TG}, we report results for density and momentum distribution functions in the Tonks-Girardeau limit obtained using the lattice approach and the worm algorithm. Given the constraint of very low site occupancies in the lattice, our lattice calculations can only be done for significantly smaller numbers of particles than within the worm algorithm. The normalized density profiles in Figs.~\ref{fig:mdistH_TG}(a) and \ref{fig:mdistH_TG}(c) exhibit the expected collapse, even for as few as nine bosons [Fig.~\ref{fig:mdistH_TG}(a)]. Data collapse can also be seen in the momentum distribution functions, except at very low momenta. The $k^{-4}$ high momentum tails are apparent both in the lattice and worm algorithm results, though they are significantly better seen in the lattice calculation where much smaller values of $m(k)$, for higher values of $k$, can be computed.

We note that, using the relations we established in Sec.~\ref{sec:hamilQMC} between quantities in the continuum and in the lattice, one finds that the scaling properties of trapped systems in the continuum discussed so far are consistent with those discussed for lattice systems in Ref.~\cite{rigolUni,rigolGS}.

\subsubsection{Finite interaction strength}

For finite interaction strengths, the relation between the chemical potential and the density is not as simple as in the Tonks-Girardeau limit, where $\mu=\hbar^2\pi^2\rho^2/(2m)$. Instead, $\mu=\hbar^2\rho^2f(\gamma)/(2m)$, where $f(\gamma)=3e(\gamma)-\gamma e'(\gamma)$ \cite{liebI}. This means that, using Eq.~\eqref{eq:eff_CP} and within LDA,
\begin{equation}
 n^2(x)f[\gamma(x)]=n^2(0)f[\gamma(0)]-\frac{x^2}{a^4_\mathrm{HO}} .
\end{equation}
The position in the trap at which the density vanishes is $L_\text{LL}=\sqrt{f[\gamma(0)]}n(0)a_\mathrm{HO}^2$. Defining, as for the Tonks-Girardeau limit, $\tilde{n}(x)=n(x)/n(0)$ and $\tilde{x}=x/L_\text{LL}$, and noticing that $\gamma(x)=mg/[\hbar^2n(x)]=\gamma(0)/\tilde{n}(x)$, we obtain
\begin{equation}
 \tilde{n}^2(\tilde{x})f\left[\frac{\gamma(0)}{\tilde{n}(\tilde{x})}\right]=f[\gamma(0)](1-\tilde{x}^2) .
 \label{eq:LDA_Gama}
\end{equation}

Equation~\eqref{eq:LDA_Gama} makes apparent that the scaled density profiles of systems with the same value of $\gamma$ in the center of the trap must be identical. As in the Tonks-Girardeau limit, this result and the LDA allow one to advance a scaling collapse in the momentum distribution of trapped systems at high momenta \cite{LLtail}. 

However, the scaling above requires the knowledge of $n(0)$ and $L_\text{LL}$, which can only be obtained numerically (or measured experimentally). We can predict another universal way to scale densities and positions by noting that, from Eq.~\eqref{eq:LDA_Gama}, it follows that
\begin{equation}\label{eq:ldallm}
 \tilde{n}(\tilde{x})=h_0[1-\tilde{x}^2,\gamma(0)],
\end{equation}
where $h_0(x,y)$ is an unknown function that can, in principle, be computed numerically (from now on we will call dummy functions of this kind as $h_\alpha$, with $\alpha=0,1,\ldots$). Since $N_b=2\int_0^{L_\text{LL}}n(x)dx=2n^2(0)a^2_\text{HO}\sqrt{f[\gamma(0)]}\int_0^1h_0[1-\tilde{x}^2,\gamma(0)]d\tilde{x}$, we see that $n(0)=h_1[\gamma(0)]\sqrt{N_b}/a_\text{HO}=h_2[\gamma(0)]\tilde{\rho}$, and $\tilde{\rho}$ is the density in the center of the trap in the Tonks-Girardeau limit. From this result for $n(0)$ it follows that $L_\text{LL}=h_3[\gamma(0)]\tilde{L}$, where $\tilde{L}$ is the point at which the density vanishes in the Tonks-Girardeau limit. Hence, after fixing $\gamma(0)$, one can use exactly the same scaling for finite interaction strengths as the one used in the Tonks-Girardeau limit. We should stress that this is unique to bosons in the continuum. In the Bose-Hubbard model, one needs two parameters (the so-called characteristic density and the on-site interaction strength) to be kept fixed in order to be able to scale density profiles \cite{rigol_batrouni_09}.

\begin{figure}[!t]
 \centering
 \includegraphics[width=1.0\linewidth]{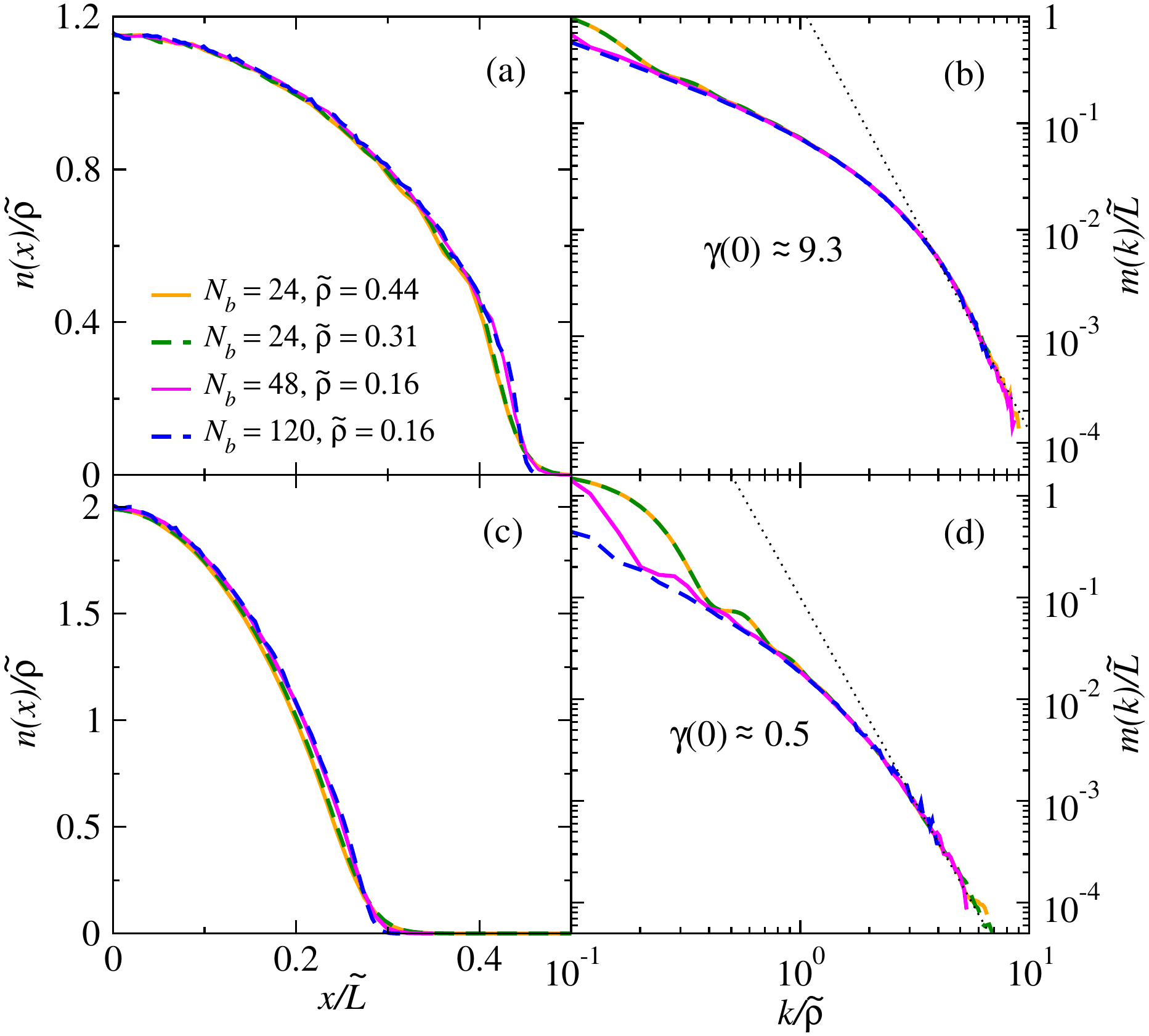}
 \caption{(Color online) [(a) and (c)] Universal behavior of scaled density profiles and [(b) and (d)] momentum distribution functions for harmonically trapped systems with $\gamma(0)\approx9.3$ in (a) and (b), and $\gamma(0)\approx0.5$ in (c) and (d). In (b) and (d), thin dotted lines depict $k^{-4}$ behavior.}
 \label{fig:mdistH_Gama}
\end{figure}

In Fig.~\ref{fig:mdistH_Gama}, we show the scaled density [Figs.~\ref{fig:mdistH_Gama}(a) and \ref{fig:mdistH_Gama}(c)] and momentum [Figs.~\ref{fig:mdistH_Gama}(b) and \ref{fig:mdistH_Gama}(d)] profiles for systems with different numbers of particles, trapping frequencies, and 1D scattering lengths. Results are presented for $\gamma(0)\approx9.3$ [Figs.~\ref{fig:mdistH_Gama}(a) and \ref{fig:mdistH_Gama}(b)] and $\gamma(0)\approx2.9$ [Figs.~\ref{fig:mdistH_Gama}(c) and \ref{fig:mdistH_Gama}(d)]. The plots exhibit an excellent data collapse in the density profiles, and in the momentum distribution functions at intermediate and high momentum, as predicted by the analysis above. The high momentum $k^{-4}$ tails are barely apparent in Figs.~\ref{fig:mdistH_Gama}(b) and \ref{fig:mdistH_Gama}(d). By comparing the results for $\gamma(0)\approx9.3$ to those in the Tonks-Girardeau limit in Fig.~\ref{fig:mdistH_TG}, one can see that, even for that large value of $\gamma(0)$, the range and sharpness of the $k^{-4}$ tails have decreased. This worsens as $\gamma(0)$ is further decreased and, as in the homogeneous case, poses an increasing experimental challenge for detecting the $k^{-4}$ tails as one departs from the Tonks-Girardeau regime.

\section{Finite Temperature}\label{sec:finitetemperature}

\subsection{Homogeneous Systems}

\subsubsection{Tonks-Girardeau limit}

\begin{figure}[!b]
 \centering
 \includegraphics[width=1.0\linewidth]{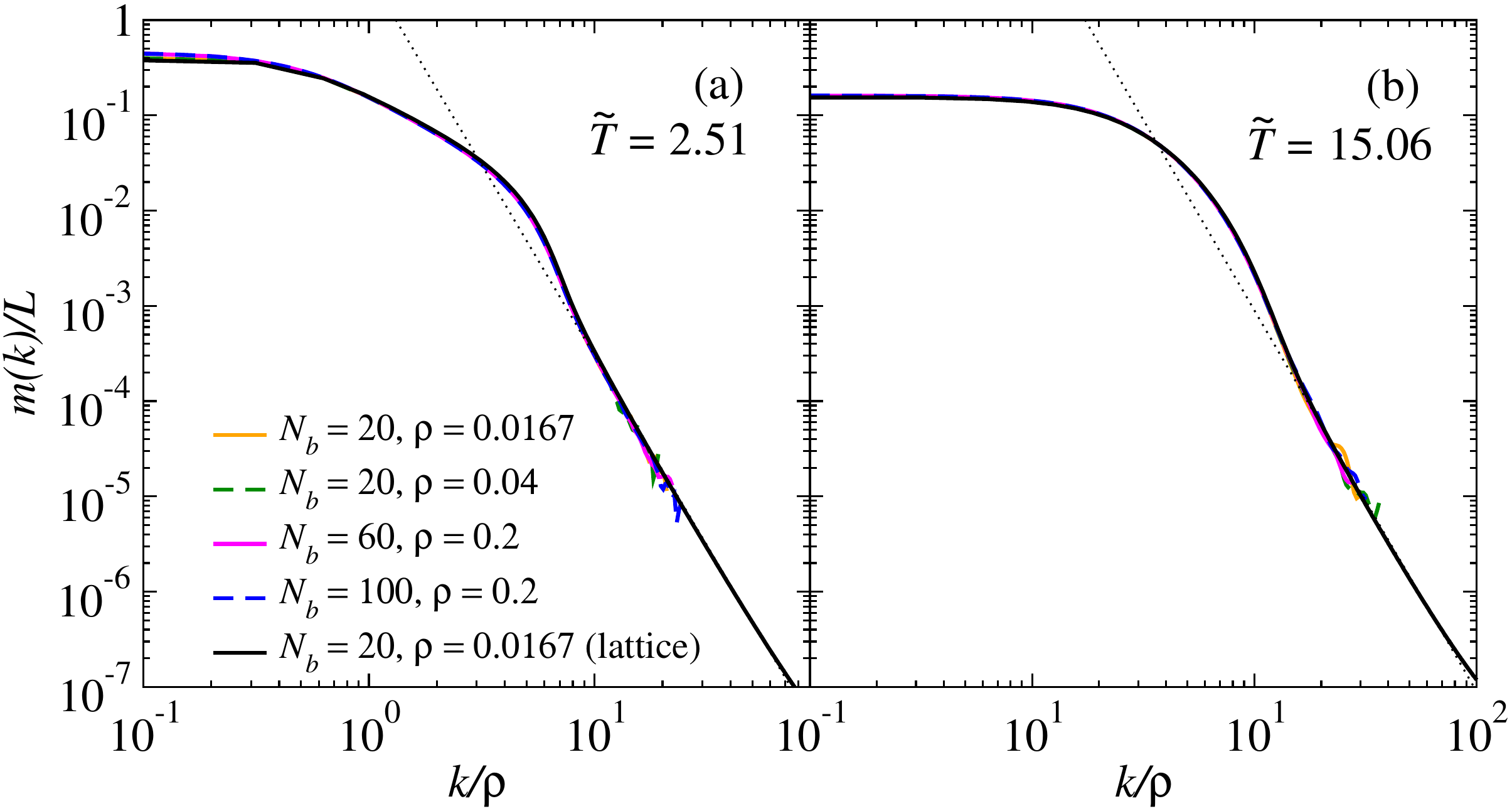}
 \caption{(Color online) Finite temperature results for the scaled momentum distribution function of the homogeneous Tonks-Girardeau gas in systems with different average number of particles and densities. The scaled temperature in the plots is: (a) $\tilde T=2.51$ and (b) $\tilde T=15.06$. The black solid line in both panels was obtained using the lattice approach, while all other curves were obtained using the worm algorithm. Thin dotted lines depict $k^{-4}$ behavior.}
 \label{fig:mdistT_TG}
\end{figure}

At finite temperature, the long-distance behavior of one-particle correlation functions is qualitatively different from the one in the ground state. This because quasi-long-range order is destroyed by thermal fluctuations, i.e., the correlation functions decay exponentially with increasing distance \cite{1Dbosonrmp}. As a result, the population of all momentum modes $m(k)/L$ becomes intensive. This means that, if the density and the temperature are kept constant, the momentum distribution function of systems with different number of particles should collapse (provided the system sizes are much larger than the correlation length). The curves with the same value of $\rho$ in Figs.~\ref{fig:mdistT_TG}(a) and \ref{fig:mdistT_TG}(b) show that this is indeed the case. 

The question we would like to address here is how to scale the momentum profiles when the density and the temperature are changed in these systems. For that, as we did in Sec.~\ref{sec:homT0TG}, we look into Tan's contact in the limit $\gamma\rightarrow\infty$. As follows from the derivation in Appendix~\ref{app:YY_bosons}, the Tan's contact takes the form~\cite{LLreview,guanpolylog,guancontact}
\begin{equation}
 \mathcal{C}=\frac{(2mk_BT)^2L}{\hbar^42\pi}f_{1/2}\left(\frac{\mu}{k_BT}\right)f_{3/2}\left(\frac{\mu}{k_BT}\right) ,
\end{equation}
where $f_{\nu}(\cdot)$ is the Fermi-Dirac function (see Appendix~\ref{app:YY_bosons} for its definition). From Eq.~\eqref{eq:fermi}, one can see that $\mu/(k_BT)$ is a function of $2mk_BT/(\hbar^2\rho^2)$. Thus, one can define the scaled (dimensionless) temperature $\tilde{T}=2mk_BT/(\hbar^2\rho^2)$. The scaled contact $\tilde{\mathcal C}=C/(L\rho^4)$ then has the form
\begin{equation}
 \tilde{\mathcal C}=\frac{\tilde{T}^2}{2\pi}f_{1/2}\left(\frac{\mu}{k_BT}\right)f_{3/2}\left(\frac{\mu}{k_BT}\right) ,
\end{equation}
which is only a function of $\tilde T$. This means that, once $\tilde{T}$ is fixed, $m(k)/L$ vs $k/\rho$ is universal. The results in Figs.~\ref{fig:mdistT_TG}(a) and \ref{fig:mdistT_TG}(b), in which we plot systems with different densities and temperatures but the same values of $\tilde{T}$, show that this is indeed the correct scaling. The $k^{-4}$ momentum tails are clearly visible at high momenta. Two important features of those tails are apparent when comparing Figs.~\ref{fig:mdistT_TG}(a) and \ref{fig:mdistT_TG}(b). The first one is that the weight of the tails (the contact) increases with increasing $\tilde{T}$, as discussed in Ref.~\cite{FTmdf}. The second one, probably more important for experiments, is that with increasing $\tilde{T}$ the values of $m(k)$ at which the tails start to develop become smaller. This means that the $k^{-4}$ momentum tails are more difficult to observe experimentally as the temperature increases, despite the fact that Tan's contact becomes larger.

\subsubsection{Finite interaction strength}

\begin{figure}[!b]
 \centering
 \includegraphics[width=1.0\linewidth]{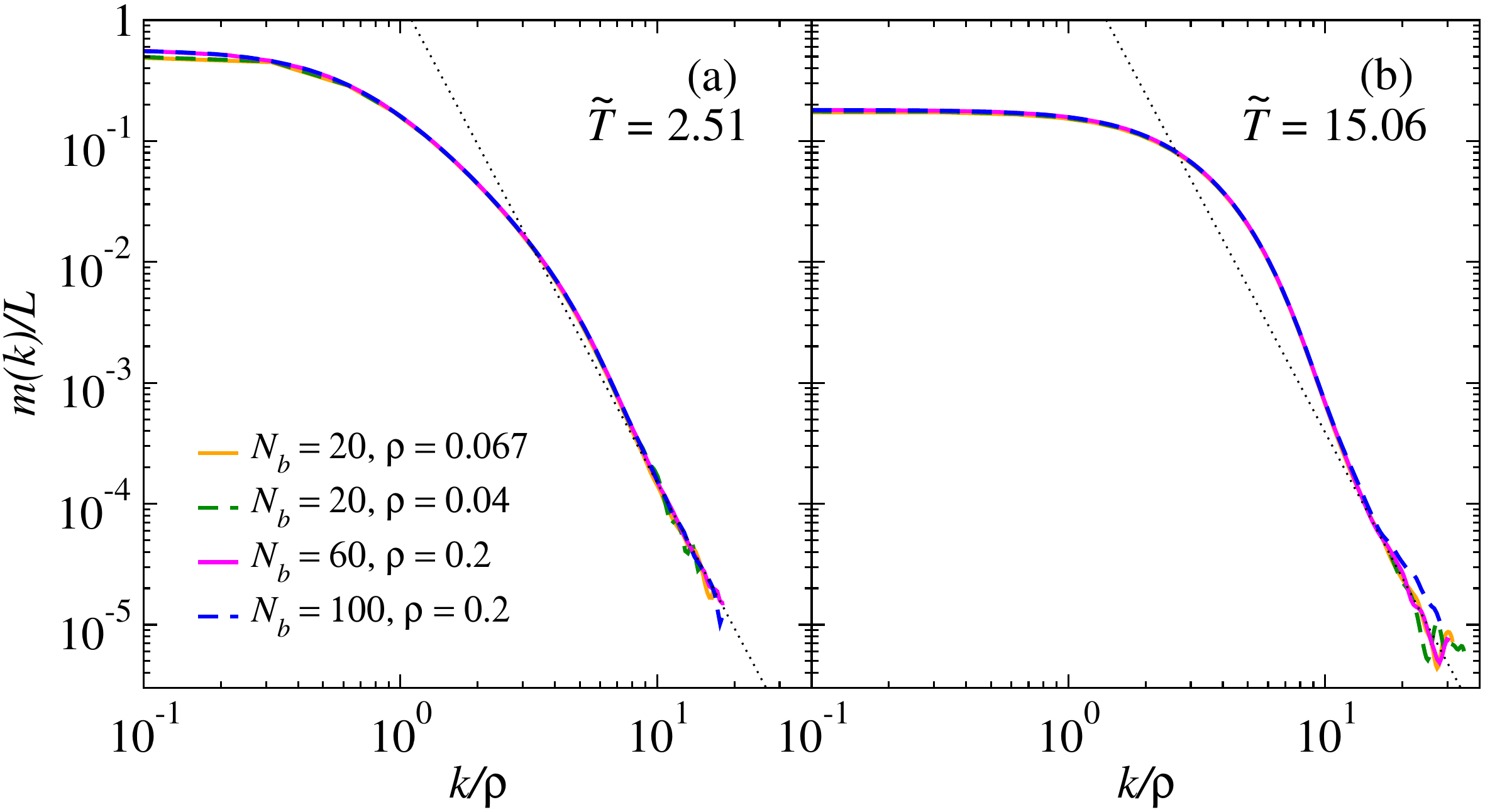}
 \caption{(Color online) Finite temperature results for the scaled momentum distribution function of the homogeneous Lieb-Liniger gas with different average number of particles and densities at finite $\gamma=8.33$. The scaled temperature in the plots is: (a) $\tilde{T}=2.51$ and (b) $\tilde T=15.06$. All results were obtained using the worm algorithm. Thin dotted lines depict $k^{-4}$ behavior.}
 \label{fig:mdistT_Gama}
\end{figure}

For systems with a finite value of $\gamma$ and finite temperatures one could advance, based on the results for the ground state (in which scaling collapse was found so long as $\gamma$ was kept fixed) and on the results for finite temperature in the Tonks-Girardeau limit (in which scaling collapse was found so long as $\tilde{T}$ was kept fixed), that scaling collapse requires us to keep $\gamma$ and $\tilde{T}$ constant. The results reported in Fig.~\ref{fig:mdistT_Gama} for a finite value of $\gamma=8.33$, which parallel the results in the Tonks-Girardeau limit reported in Fig.~\ref{fig:mdistT_TG}, show that this is indeed the case. The behavior of the high momentum tails with increasing $\tilde{T}$ in Fig.~\ref{fig:mdistT_Gama} is qualitatively similar to that discussed in the Tonks-Girardeau limit.

While a close expression for Tan's contact for finite values of $\gamma$ and finite temperatures is not available,
a calculation up to order $1/\gamma$ (see Appendix~\ref{app:YY_bosons}) reveals that $\tilde{\mathcal C}=C/(L\rho^4)$ can be written as~\cite{LLreview,guanpolylog,guancontact}
\begin{equation}
 \tilde{\mathcal C}=\frac{\tilde{T}^2}{2\pi}f_{3/2}\times\left[f_{1/2}+\frac{\tilde{T}^{1/2}}{\sqrt\pi\gamma}(2f^2_{1/2}+f_{-1/2}f_{3/2})\right] ,
\end{equation}
where by $f_{\nu}$ it is meant $f_{\nu}(\mu/k_BT)$. This further supports the correctness of the scaling used in Fig.~\ref{fig:mdistT_Gama}.

\subsection{Trapped Systems}

\subsubsection{Tonks-Girardeau limit}

\begin{figure}[!b]
 \centering
 \includegraphics[width=1.0\linewidth]{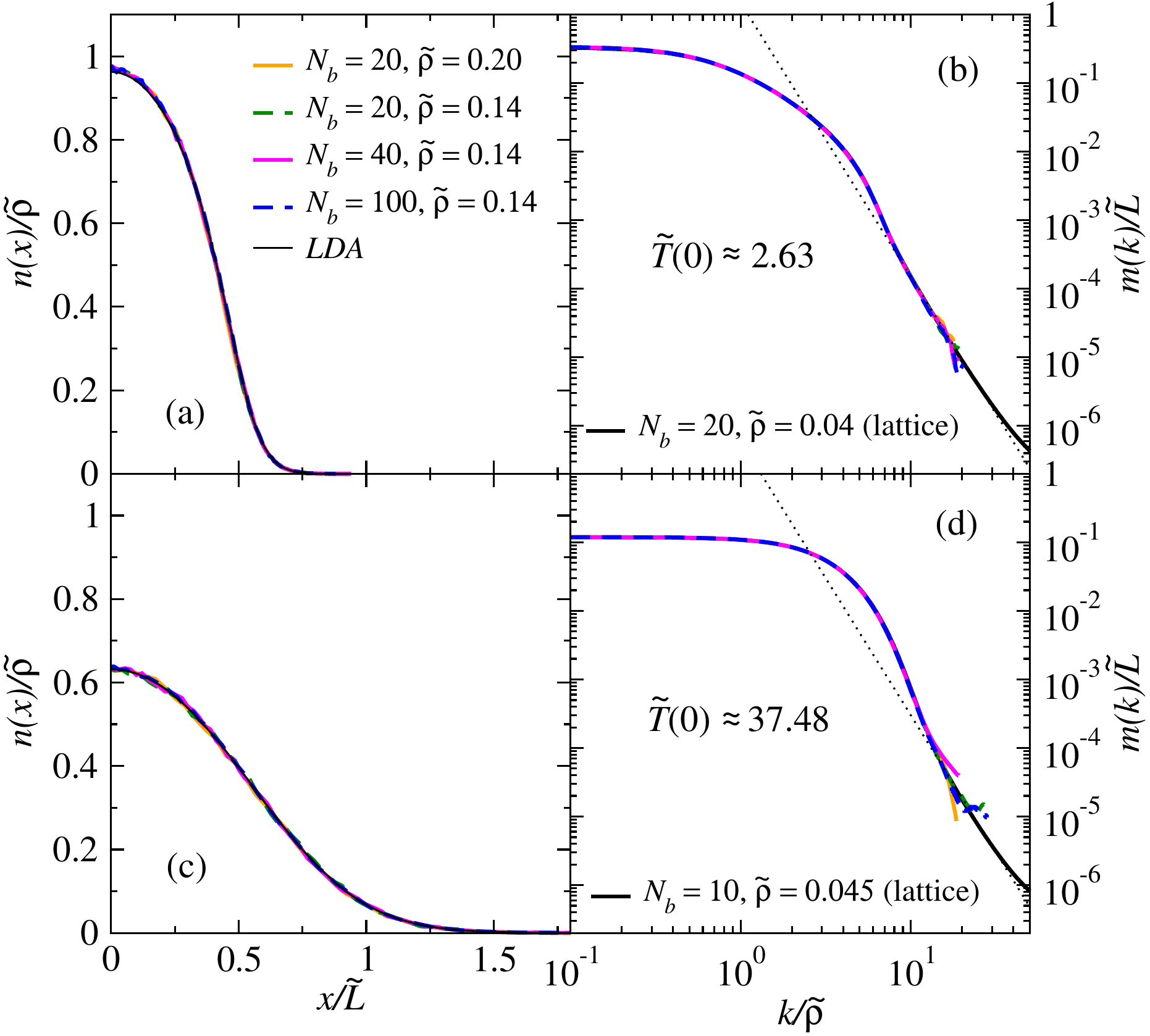}
 \caption{(Color online) Finite temperature results for [(a) and (c)] the scaled density profiles and [(b) and (d)] momentum distribution functions of harmonically trapped Tonks-Girardeau gases with [(a) and (b)] $\tilde{T}(0)\approx2.63$ and [(c) and (d)] $\tilde{T}(0)\approx37.48$. In (a) and (c), black solid lines show the LDA predictions. In (b) and (d), the black solid lines show results of the lattice calculations. All other curves were obtained using the worm algorithm. In (b) and (d), thin dotted lines depict $k^{-4}$ behavior.}
 \label{fig:mdistHT_TG}
\end{figure}

As for the ground-state case, let us discuss how to scale density profiles in the Tonks-Girardeau limit at finite temperature. Dividing Eq.~\eqref{eq:eff_CP} by $k_BT$, we obtain
\begin{equation}
 \frac{\mu(x)}{k_BT}=\frac{\mu(0)}{k_BT}-\frac{m\omega^2x^2}{2k_BT},
 \label{eq:eff_CPT}
\end{equation}
Using the LDA and the fact that, as discussed in the previous section, in homogeneous systems $\mu/(k_BT)=h_4(2mk_BT/\hbar^2\rho^2)$, we can rewrite Eq.~\eqref{eq:eff_CPT} as
\begin{equation}
 h_4\left[\frac{\tilde{T}(0)n^2(0)}{n^2(x)}\right]=h_4[\tilde{T}(0)]-\frac{x^2}{a_{HO}^4\tilde{T}(0)n^2(0)},
 \label{eq:eff_CPT1}
\end{equation}
where $\tilde{T}(0)=2mk_BT/[\hbar^2n^2(0)]$. Now, let us find the position $L_\text{FT}$ in the trap at which the chemical potential vanishes. At zero temperature, the position at which the chemical potential vanishes is also the position at which the density vanishes. However, this is not the case anymore at finite temperature. For $\mu(x)=0$, $h_4[2mk_BT/\hbar^2n^2(x)]=0$ and, using Eq.~\eqref{eq:eff_CPT1}, $L_\text{FT}=\sqrt{\tilde{T}(0)h_4[\tilde{T}(0)]}n(0)a_{HO}^2$. Defining $\tilde{x}=x/L_\text{FT}$, Eq.~\eqref{eq:eff_CPT1} can be rewritten as
\begin{equation}
 h_4\left[\frac{\tilde{T}(0)n(0)^2}{n(x)^2}\right]=h_4[\tilde{T}(0)](1-\tilde{x}^2).
 \label{eq:eff_CPT2}
\end{equation}
Defining $\tilde{n}(x)=n(x)/n(0)$, we see that Eq.~\eqref{eq:eff_CPT2} implies that
\begin{equation}
 \tilde{n}(x)=h_5[1-\tilde{x}^2,\tilde{T}(0)].
 \label{eq:eff_CPT3}
\end{equation}
At this point, we note that Eq.~\eqref{eq:eff_CPT3} has exactly the same form as Eq.~\eqref{eq:ldallm}, with $h_0\rightarrow h_5$ and $\gamma(0)\rightarrow\tilde{T}(0)$. Using that $N_b=2\int_0^{\infty}n(x)dx$, we get that $n(0)=h_6[\tilde{T}(0)]\tilde{\rho}$ and $L_\text{FT}=h_7[\tilde{T}(0)]\tilde{L}$. This means that, once $\tilde{T}(0)$ has been fixed, a scaling as the one used for the ground state case will produce universal density profiles at finite temperature. In other words, for a fixed $\tilde{T}(0)$, $\tilde{\rho}$ and $\tilde{L}$ determine the average density in the system and an effective system size. Hence, given the fact that LDA is expected to work for sufficiently large system sizes at any finite temperature (because of finite correlation lengths), plots of $m(k)/\tilde{L}$ vs $k/\tilde{\rho}$ are expected to exhibit data collapse.

In Figs.~\ref{fig:mdistHT_TG}(a) and \ref{fig:mdistHT_TG}(c), we show plots of $n(x)/\tilde{\rho}$ vs $x/\tilde{L}$ for systems with different average number of particles, densities in the center of the trap, and temperatures, but all with the same value of $\tilde{T}(0)\approx2.63$ [Fig.~\ref{fig:mdistHT_TG}(a)] and $\tilde{T}(0)\approx37.48$ [Fig.~\ref{fig:mdistHT_TG}(c)]. All the scaled density profiles exhibit the predicted data collapse and are in excellent agreement with the LDA prediction. In Figs.~\ref{fig:mdistHT_TG}(b) and \ref{fig:mdistHT_TG}(d), we show the corresponding scaled momentum distribution functions, which also exhibit an almost perfect data collapse. The results for the momentum distribution functions exhibit an excellent agreement between the worm algorithm calculations and the lattice approach results. The latter allow us to reach smaller values of $m(k)/\tilde{L}$ (larger values of $k/\tilde{\rho}$) so that the $k^{-4}$ momentum tails can be better resolved.

\subsubsection{Finite interaction strength}

\begin{figure}[!t]
 \centering
 \includegraphics[width=1.0\linewidth]{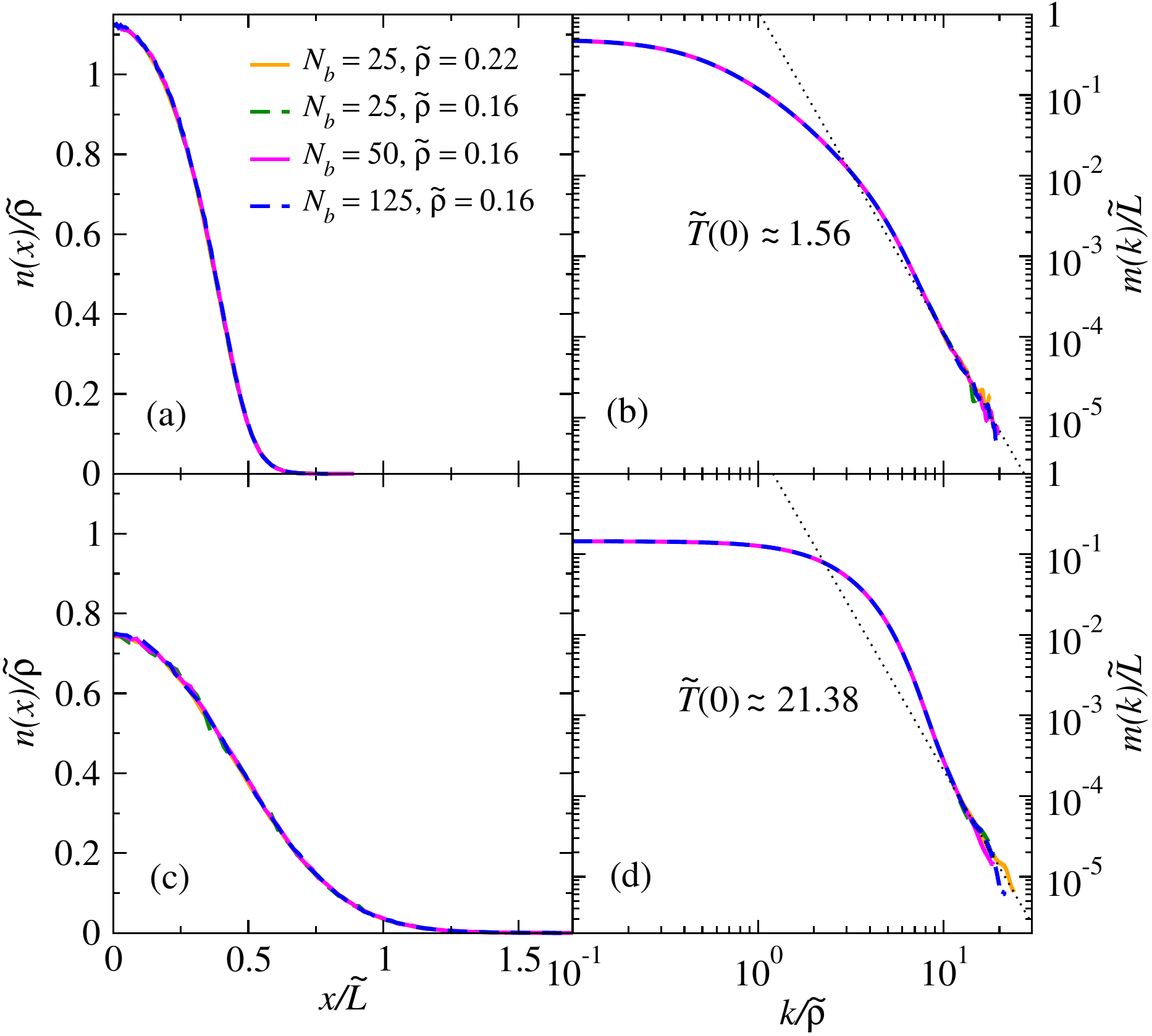}
 \caption{(Color online) Finite temperature results for [(a) and (c)] the scaled density profiles and [(b) and (d)] momentum distribution functions of harmonically trapped Lieb-Liniger gases with [(a) and (b)] $\gamma(0)\approx9.3$, and $\tilde{T}(0)\approx1.56$ and [(c) and (d)] $\tilde{T}(0)\approx21.38$. All curves were obtained using the worm algorithm. In (b) and (d), thin dotted lines depict $k^{-4}$.}
 \label{fig:mdistHT_Gama}
\end{figure}

The scaling of trapped systems with finite interaction strengths at finite temperature is the most challenging one. From the ground state analysis in the presence of a trap we know that $\gamma(0)$ needs to be kept fixed. From the finite temperature analysis of homogeneous systems we learned that $\gamma$ and $\tilde{T}$ need to be fixed, while the finite temperature analysis of the trapped Tonks-Girardeau gas revealed that $\tilde{T}(0)$ needs to be fixed. Putting all this together we can advance that, for finite interaction strength, trapped systems at finite temperature will only exhibit scaling collapse if $\gamma(0)$ and $\tilde{T}(0)$ are fixed. We can also advance, from the analysis of trapped systems with finite interaction strength in the ground state and in the Tonks-Girardeau limit at finite temperature that using $\tilde{\rho}$ and $\tilde{L}$ will allow one to achieve the expected collapse.

In Fig.~\ref{fig:mdistHT_Gama} we show results for the scaled momentum distribution function of trapped systems with the same value of $\gamma(0)$ and two values of $\tilde{T}(0)$. They all exhibit a nearly perfect collapse. The features observed in the momentum tails of the homogeneous case are also apparent here. Namely, Tan's contact increases with increasing temperature but at the same time the $k^{-4}$ tails develop starting at higher values of $k$  and smaller values of $m(k)$.

\section{Conclusions}\label{sec:conclusion}

We carried out an unbiased study of one-dimensional trapped bosons in the ground state and at finite temperature using the worm algorithm, and the Bose-Fermi mapping in the presence of a lattice at low fillings. Whenever possible, we compared our results to the predictions of the Bethe ansatz obtaining an excellent agreement.

Our study focused in the behavior of the density and momentum distribution functions, as well as one-particle correlations. We discussed in detail how to scale density and momentum profiles to observe universal behavior, which was demonstrated by our numerical results. For trapped systems in the ground state, we showed that fixing $\gamma$ in the center of the trap is all that is required to obtain a universal scaling of density and momentum profiles. This is to be contrasted with the Bose-Hubbard model, for which one needs to fix the so-called characteristic density and the on-site interaction strength \cite{rigol_batrouni_09}. At finite temperature in systems in the continuum, in addition to the condition for the ground state, one needs to fix the scaled temperature in the center of the trap. Those two parameters fully characterize the densities and momentum distribution functions. 

In our study of the momentum distribution function, we payed special attention to the $k^{-4}$ asymptotic behavior at high momenta. As shown by our numerical calculations, those momentum tails become increasingly challenging to resolve as the temperature increases and the interactions become weaker, making the Tonks-Girardeau limit at low temperatures the ideal regime for their experimental observation.

\begin{acknowledgments}
This work was supported by the National Science Foundation Grant No.~PHY13-18303 and by the Office of Naval Research. The computations were performed in the Institute for CyberScience at Penn State and the Center for High-Performance Computing at the University of Southern California. We thank Nikolay Prokof'ev for providing us with the worm algorithm code used in the calculations, Jean-S\'ebastien Caux for providing us with the Bethe ansatz results reported in Fig.~\ref{fig:LL_JCaux}, and Juan Carrasquilla and David Weiss for insightful discussions.
\end{acknowledgments}

\newpage\

\bibliography{mdist}

\appendix

\section{Pair-product action} \label{app:PPU}
The pair-product action assumes that interactions between different pairs of particles are uncorrelated. Thus, the action from the interaction breaks into sum of pair actions
\begin{equation}
 U_2(R,R';\tau)=\sum_{i<j}u_2(r_{ij},r'_{ij};\tau) .
\end{equation}
The pair action for particles with repulsive contact interactions can be obtained exactly from the relative two-body propagator $\rho_\delta$, $u_2=-\ln\rho_\delta$. The relative two-body propagator is the two-body propagator divided by the free propagator $\rho_\delta=G_2/G_0$. For the Lieb-Liniger Hamiltonian, written as
\begin{equation}
\hat H=-\lambda\sum_{i=1}^N\partial_{x_i}^2+g\sum_{i<j=1}^{N}\delta(x_i-x_j),
\end{equation}
the relative propagator can be written as
\begin{equation}
\begin{split}
&\rho_{\delta}(x,y;\tau)=1-\frac{g}{2}\sqrt\frac{\pi\tau}{\lambda'}\times\mathrm{erfc}\left[\frac{|x|+|y|+g\tau}{\sqrt{4\lambda'\tau}}\right] \\ &\times\exp\left[\frac{1}{4\lambda'\tau}(x-y)^2+\frac{g}{2\lambda'}(|x|+|y|)+\frac{\tau g^2}{4\lambda'}\right] ,
\label{eq:propagator}
\end{split}
\end{equation}
where $\lambda'=2\lambda$ and $\mathrm{erfc}[x]=2/\sqrt\pi\int_x^{\infty}dt\exp(-t^2)$ is the complementary error function (see Refs.~\cite{manoukian_edward_89,casula_ceperley_08} for the analytic derivation of this result).

\section{Energy estimators} \label{app:E_est}

The thermodynamic estimator for the total energy is computed from $E_{T}=-Z^{-1}dZ/d\beta$, while the one for the kinetic energy is computed from $K_T=-m\beta^{-1}Z^{-1}dZ/dm$~\cite{pimc,worm2}. One obtains that
\begin{equation}
\begin{split}
&E_T=\left\langle\frac{1}{2\tau}-\frac{(R_{\alpha}^i-R_{\alpha}^{i-1})^2}{4\lambda\tau^2}+
 \frac{dU_{\alpha}^i}{d\tau}\right\rangle_{\alpha,i} \\
&K_T=\left\langle\frac{1}{2\tau}-\frac{(R_{\alpha}^i-R_{\alpha}^{i-1})^2}{4\lambda\tau^2}+\frac{\lambda}{\tau}
 \frac{dU_{\alpha}^i}{d\lambda}\right\rangle_{\alpha,i} \\
&V_T=\left\langle\frac{dU_{\alpha}^i}{d\tau}-\frac{\lambda}{\tau}\frac{dU_{\alpha}^i}{d\lambda}\right\rangle_{\alpha,i} .
\end{split}
\end{equation}

Here, $i$ and $\alpha$ are indices for particle and discrete imaginary time, respectively. $U_{\alpha}^i$ and its derivatives can be computed using the pair-product action discussed in Appendix~\ref{app:PPU}.

\section{Yang-Yang Thermodynamics} \label{app:YY_bosons}
The Yang-Yang equation for the dressed energy is \cite{guanpolylog}
\begin{equation}
\begin{split}
\epsilon(k)=&\frac{\hbar^2k^2}{2m}-\mu \\
&-\frac{k_BT}{2\pi}\int_{-\infty}^{\infty}dq\frac{2c}{c^2+(k-q)^2}\ln[1+e^{-\frac{\epsilon(q)}{k_BT}}].
\end{split}
\end{equation}
$\epsilon(k)$ is used to compute thermodynamic quantities. The thermodynamic potential density is given by the expression
\begin{equation}
 \Omega(\mu,c,T)=-\frac{k_BT}{2\pi}\int_{-\infty}^{\infty}dq\ln[1+e^{-\frac{\epsilon(q)}{k_BT}}].
\end{equation}
All other thermodynamic quantities of interest here can be derived from $\Omega(\mu,c,T)$ through the following relations
\begin{equation}
 e=\Omega+\mu\rho+Ts,\;p=-\Omega,\;\rho=-\frac{\partial \Omega}{\partial\mu},\;s=-\frac{\partial \Omega}{\partial T},
 \label{eq:tq}
\end{equation}
where $e$ is the energy density, $s$ is the entropy density, and $p$ is the pressure.

For very strong contact interactions, i.e., $c\gg1$, one can Taylor expand $\epsilon(k)$ in terms of $1/c$
\begin{equation}
 \epsilon(k)=\frac{\hbar^2k^2}{2m}-\mu+\frac{2}{c}\Omega+O\left(\frac{1}{c^3}\right).
 \label{eq:taylor}
\end{equation}
The dressed energy $\epsilon(k)$ can then be obtained iteratively. To zeroth order: $\epsilon^{(0)}(k)=\hbar^2k^2/(2m)-\mu$. This allows one to compute $\Omega$ to lowest order
\begin{equation}
\Omega^{(0)}=-\frac{\sqrt{m}(k_BT)^{3/2}}{\hbar\sqrt{2\pi}}f_{3/2}\left(\frac{\mu}{k_BT}\right),
\label{eq:pt0}
\end{equation}
where \begin{equation}
 f_\nu(\alpha)=\frac1{\Gamma(\nu)}\int_0^\infty\frac{x^{\nu-1}dx}{e^{(x-\alpha)}+1} .
\end{equation}
is the Fermi-Dirac function. Using Eq.~\eqref{eq:pt0}, and the relations \eqref{eq:tq}, one obtains the following expressions for the density and the energy of the system
\begin{subequations}
\begin{align}
&\rho^{(0)}=\frac{\sqrt{m}(k_BT)^{1/2}}{\hbar\sqrt{2\pi}}f_{1/2}\left(\frac{\mu}{k_BT}\right),\\
&e^{(0)}=\frac{\sqrt{m}(k_BT)^{3/2}}{2\hbar\sqrt{2\pi}}f_{3/2}\left(\frac{\mu}{k_BT}\right).
\end{align}
\label{eq:fermi}
\end{subequations}
These are nothing but the density and the energy density of a system of noninteracting spinless fermions.

Substituting $\Omega\rightarrow\Omega^{(0)}$ in Eq.~\eqref{eq:taylor}, one obtains
\begin{eqnarray}
\epsilon^{(1)}(k)&=&k^2-\mu^{(1)}, \quad\text{where} \\
\mu^{(1)}&=&\mu+\frac{\sqrt{2m}(k_BT)^{3/2}}{\hbar\sqrt\pi c}f_{3/2}\left(\frac{\mu}{k_BT}\right).\nonumber
\label{eq:mu1}
\end{eqnarray}
Now, substituting $\mu\rightarrow\mu^{(1)}$ in Eq.~\eqref{eq:pt0}, and expanding to first order in $1/c$, we get
\begin{equation}
 \Omega^{(1)}=-\frac{\sqrt{m}(k_BT)^{3/2}}{\hbar\sqrt{2\pi}}f_{3/2}\times\left(1+\frac{\sqrt{2mk_BT}}{c\hbar\sqrt{\pi}}f_{1/2}\right),
\label{eq:ptaylor1}
\end{equation}
where, from now on, by $f_{\nu}$ it is meant $f_{\nu}(\mu/k_BT)$. Finally, substituting $\Omega\rightarrow\Omega^{(1)}$ in 
Eq.~\eqref{eq:taylor}, we get that
\begin{eqnarray}
\epsilon^{(2)}(k)&=&k^2-\mu^{(2)}, \quad\text{where} \\
\mu^{(2)}&=&\mu+\frac{\sqrt{2m}(k_BT)^{3/2}}{\hbar\sqrt\pi c}f_{3/2}\nonumber\\&&\times\left(1+\frac{\sqrt{2mk_BT}}{c\hbar\sqrt{\pi}}f_{1/2}\right).\nonumber
\label{eq:mu2}
\end{eqnarray}
This expression allows us to obtain $\Omega^{(2)}$ by substituting $\mu\rightarrow\mu^{(2)}$ in Eq.~\eqref{eq:pt0}. Expanding the resulting equation up to second order in $1/c$, one obtains
\begin{eqnarray}
 \Omega^{(2)}&=&\frac{\sqrt{m}(k_BT)^{3/2}}{\hbar\sqrt{2\pi}}f_{3/2}\times\left[1+\frac{\sqrt{2mk_BT}}{c\hbar\sqrt{\pi}}f_{1/2}\right.\nonumber\\
 &&+\left.\left(\frac{\sqrt{2mk_BT}}{c\hbar\sqrt{\pi}}\right)^2\left(f^2_{1/2}+\frac{1}{2}f_{3/2}f_{-1/2}\right)\right].\qquad
\label{eq:ptaylor}
\end{eqnarray}

Using Eq.~\eqref{eq:ptaylor}, we can calculate Tan's contact through Tan's sweep relation
\begin{equation}
 \mathcal{C}=-\frac{2mc^2L}{\hbar^2}\left(\frac{\partial p}{\partial c}\right)_{\mu,T} .
\end{equation}
The result obtained to order $1/c$ is:
\begin{eqnarray}\label{eq:YYcontact}
\mathcal{C}&=&\frac{(2mk_BT)^2L}{\hbar^42\pi}f_{3/2} \\
&&\times\left[f_{1/2}+\frac{\sqrt{2m}(k_BT)^{1/2}}{\hbar\sqrt\pi c}(2f^2_{1/2}+f_{-1/2}f_{3/2})\right] .
\nonumber 
\end{eqnarray}

\clearpage
\end{document}